\documentclass[twocolumn,showpacs,preprintnumbers,amsmath,amssymb,aps]{revtex4-2}

\usepackage{graphicx}
\usepackage{eurosym}
\usepackage[T1]{fontenc}
\usepackage[usenames,dvipsnames,table]{xcolor}
\usepackage[utf8]{inputenc}
\usepackage{natbib}
\usepackage{lmodern}
\usepackage{booktabs}
\usepackage{amsmath,amssymb,amsfonts,latexsym}
\usepackage[colorlinks=true,linktocpage=true,linkcolor=blue,citecolor=blue]{hyperref}
\usepackage{xcolor}
\usepackage{ulem} 

\def\sst{\scriptscriptstyle}
\newcommand{\eqn}[1]{Eq.~\eqref{#1}}

\long\def\comment#1{ }

\def\0{{\boldsymbol 0}}

\def\and{\qquad\text{and}\qquad}

\def\med{\text{med}}

\def\min{\text{min}}
\def\max{{\text{max}}}

\def\pT{p_{\sst T}}
\def\pTj{p_{\sst T}^\text{jet}}
\def\pTg{p_{\sst T}^\gamma}
\def\xjg{x_{\text{J}\gamma}}

\renewcommand{\emph}{\textit}

\newcommand{\beq}{\begin{eqnarray}}
\newcommand{\eeq}{\end{eqnarray}}
\newcommand{\be}{\begin{eqnarray*}}
\newcommand{\ee}{\end{eqnarray*}}
\newcommand{\bal}{\begin{align}}
\newcommand{\eal}{\end{align}}
\newcommand{\rmd}{{\rm d}}
\newcommand{\dd}{{\rm d}}
\newcommand{\rme}{{\rm e}}

\begin{document}

\title{Constraining Jet Quenching in Heavy-Ion Collisions with Bayesian Inference}

\author{Alexandre Falcão}
 \email{alexandre.falcao@uib.no}
\author{Konrad Tywoniuk}
 \email{konrad.tywoniuk@uib.no}
\affiliation{Department of Physics and Technology, University of Bergen, 5007 Bergen, Norway}

\date{\today}

\begin{abstract}
Jet suppression and modification is a hallmark feature of heavy-ion collisions. This can be attributed to an accumulated set of effects, including radiative and elastic energy loss and reabsorption of thermalized energy within the jet cone, which are encoded in a \textit{quenching weight}, determining the probability distribution for a shift of the $\pT$ (energy loss). We perform a data-driven analysis, based on Bayesian inference, to extract information about the energy-loss distribution experienced by propagating jets using generic and flexible parametrizations. We first establish the consistency between different data-sets and, thereby, provide evidence for the universality of the quark/gluon quenching weights for different observables. Furthermore, we extract that the color dependence of energy loss is slightly bigger than what expected from Casimir scaling, pointing to the importance of multi-parton quenching within high-$\pT$ jets at the LHC.
\end{abstract}

\maketitle

\section{Introduction}
\label{sec:intro}

The properties of nuclear matter at high energy and matter densities have wide applications to both cosmological and astrophysical problems. Heavy-ion collisions at the LHC provide sufficient energy per unit volume to surpass the ordinary nuclear density by a factor of 20 \cite{Busza:2018rrf}. There is broad experimental and theoretical consensus that, under these conditions, the constituents of composite hadrons form a strongly-interacting, hot and dense nuclear medium, dubbed the quark-gluon plasma (QGP). This elusive state of matter leaves a characteristic imprint on a number of observables, perhaps most notably on multiparticle flow correlations, which allows to extract non-equilibrium transport coefficients to high precision.

The large colliding energies of the nucleons provide, at the same time, conditions where large transverse-momentum exchange processes happen abundantly. These so called ``hard'' processes give rise to the creation of high-$\pT$ QCD jet and heavy particle observables that are systematically described by perturbative QCD and that have often been measured to high precision in colliding systems that do not involve the formation of a QGP. The kinematic conditions at the LHC therefore provide a unique opportunity to study how perturbative QCD behaves in the background of a QGP \cite{Apolinario:2022vzg, Cunqueiro:2021wls}.

QCD jets are usually considered well-calibrated probes of the QGP, see \cite{Cunqueiro:2023vxl} for a detailed discussion. But to take a specific example, on the theory side, a first-principle description of jet evolution in matter is a daunting and still largely unresolved challenge. Different computational paradigms invoke a weakly-coupled (or ``perturbative''), such as the BDMPS-Z theory of radiative energy loss, or strongly-coupled, such as the AdS/CFT correspondence, description of the interactions between a jet and a QGP. Furthermore, since both the production of the hard probe as well as the the local conditions of the QGP it traverses are highly stochastic and fluctuating, one often have to rely on Monte Carlo approaches and modeling of various aspects of the collision.

It is therefore remarkable that, despite differences in the details of modeling various pieces of the problem, both semi-analytical approaches and sophisticated numerical frameworks are able to describe the basic observables. This raises the question of which common features among the models primarily drive the description of experimental data, and which details have little impact on the final results. One possible answer is that the models are sufficiently flexible, with enough parameters to fit the existing data within experimental uncertainties. However, it could also be that the conventionally designed observables are not sensitive to the finer details of the underlying physics. A strong argument in favor of this is the inherent bias in jet observables and other similar measures involving a hard momentum scale. Since these observables are rare probes, described by a steeply falling spectrum in transverse momentum $\pT$ or transverse mass $m_T = \sqrt{m^2+\pT^2}$, even small modifications to their kinematics can significantly alter the shape of the spectrum. As a result, the experimental samples are dominated by jets that are minimally modified\,\textemdash\, an example of the so-called ``survival bias,'' perhaps for the first time discussed in the context of heavy-ion collisions in \cite{Baier:2001yt} and studied further in \cite{Rajagopal:2016uip, Casalderrey-Solana:2018wrw}. This is for example known to prominently affect the nuclear modification factor $R_{AA}$ which measures the $\pT$ spectrum of jets produced in heavy-ion collisions compared to the equivalent in proton-proton collisions \cite{Baier:2001yt}. There have been many efforts to mitigate these bias effects using machine learning techniques \cite{Du:2020pmp,Du:2021pqa} or by defining observables with smaller sensitivity to the initial production spectrum, see, e.g. \cite{Brewer:2018dfs,Takacs:2021bpv,Apolinario:2024apr}.

Another interesting class of observables involves coincidence measurements of leading and subleading jets in opposite hemispheres of the azimuthal angle, quantified for example by the ratio of their transverse momenta. The bias effects mentioned earlier can be mitigated by considering the coincidence measurement of a high-$\pT$ jet with a back-to-back electroweak boson. Since their transverse momenta are balanced at leading order in perturbative QCD, and the electroweak boson does not interact with the QGP, its final $\pT$ serves as a reliable proxy for the initial jet $\pT$\,\textemdash\, that is, the transverse momentum of the parton before it began fragmenting and interacting with the QGP. However, higher-order contributions introduce significant deviations from this idealized back-to-back configuration, leading to a broad distribution. This, in turn, reintroduces sensitivity to the underlying spectrum and exacerbates the bias effects.

Given the rich experimental data on high-$\pT$ hadrons and fully reconstructed jets, an extensive effort has been dedicated to extract the information about the medium, most succinctly encoded in the so-called jet transport coefficient $\hat q$. Being a measure of the amount of momentum transferred from the medium to the jet, inducing both elastic and inelastic scattering processes, it is directly proportional to (a power of) the energy density of the underlying medium \footnote{At leading order, the momentum transfer is purely transverse to the direction of the jet particles and scales as $\hat q \propto \epsilon^{3/4}$, where $\epsilon$ is the energy density of the medium.}.
Model extractions of $\hat q$ was performed early comparing multiple models \cite{JET:2013cls}. A global analysis of different types of jet quenching data, including single-inclusive hadron production, di-hadron production and $\gamma$-hadron production was considered in \cite{Xie:2022ght,Xie:2022fak}, where the parametrization of temperature-dependent jet transport parameter $\hat q$ was left unconstrained. A strong temperature-dependence was obtained in the low-$T$ region, which has recently also been confirmed within a Bayesian analysis of the JETSCAPE framework that has also been extended to account for a wider class of observables, including jet spectra and substructure \cite{JETSCAPE:2024cqe}.  Despite sophisticated modeling and statistical analysis, the resulting uncertainties of $\hat q$ at both high and low temperatures remains about a factor 2--3 \cite{Xie:2022ght}.

For the last decade, Bayesian inference has been increasingly used in the context of heavy-ion collisions. As referred above, global Bayesian analyses on a wide class of observables have been used to calibrate jet energy loss models, and in that way to extract medium properties \cite{Xu:2017obm, He:2018gks, Ke:2018tsh, Ke:2020clc, JETSCAPE:2020mzn, JETSCAPE:2021ehl, JETSCAPE:2021ehl, Xie:2022fak, Xie:2022ght, Wu:2023azi, Zhang:2023oid, JETSCAPE:2023ikg, JETSCAPE:2024cqe}. Other works have focused on the extraction the shear and bulk viscosity of the QGP \cite{Novak:2013bqa, Bernhard:2015hxa, Bernhard:2016tnd, Bernhard:2018hnz, Bernhard:2019bmu, JETSCAPE:2020shq, Auvinen:2017fjw, Xie:2022fak, Nijs:2020ors}. Finally, Bayesian inference has also been used more broadly in other areas of heavy-ion collisions, for example in the constraining of the QGP equation of state \cite{Pratt:2015zsa}, or in the reweighting of nuclear PDFs \cite{Armesto:2013kqa}. More on the use of Bayesian inference for heavy-ion collisions, and specifically in jet quenching phenomenology, can be found in, e.g. \cite{Paquet:2023rfd} and \cite{Cao:2024pxc}, respectively.

In this work, we aim to determine what the data can unambiguously tell us about QCD jet quenching. The motivation is two-fold. First, we wish to assess the sensitivity of existing data, along with its associated uncertainties, to the finer details of modeling. For example, can the inclusive jet spectrum at mid-rapidity distinguish between two models of energy loss that, on average, result in a similar $\pT$ shift? While the answer might seem trivial for a single observable, we aim to quantify this by analyzing a diverse set of data sets. By `different', we do not necessarily mean mutually exclusive, but rather a collection of data points that allows us to disentangle the quark and gluon contributions to jet quenching. Second, after establishing robustness in describing the experimental data, we seek to extract generic features, independent of the specific models used, that give rise to observable trends. In our case, one such feature will be the color dependence of jet quenching, namely, the stronger suppression of gluon-initiated jets compared to quark-initiated ones \cite{Spousta:2015fca,Spousta:2016agr,Qiu:2019sfj} \footnote{We note that Refs.~\cite{Apolinario:2020nyw,Apolinario:2024apr} found a similar quenching pattern for quark- and gluon-initiated jets within the JEWEL Monte-Carlo model.}. Indeed, the importance of such separation has led to efforts to separate quark-gluon-initiated jets at the measurement level \cite{Brewer:2020och}.

Under the usual assumptions underlying QCD factorization, considering jet energy scales $\mathcal{O}(\pT\sim 10^2 \text{ GeV})$ that are much larger than the medium scales $\mathcal{O}(T \sim 0.5 \text{ GeV})$, we separate the short-distance processes creating the hard probe and its subsequent long-distance modification, see also \cite{He:2018gks}. Our starting point is therefore a factorization of the jet spectrum $\sigma^\text{med}$ in heavy-ion collisions \cite{Baier:2001yt,Qiu:2019sfj,He:2018gks}, that can be written as
\begin{equation}
    \label{eq:fact}
    \sigma^\text{med}(\pT)= D(\varepsilon)\otimes\sigma^\text{vac}(\pT+\varepsilon)\,,
\end{equation}
where $\sigma^\text{vac}$ is the jet spectrum for the equivalent kinematic process in ``vacuum'', i.e. in the absence of a medium. Here, a  jet is taken as a vacuum jet that loses an amount of energy $\varepsilon$ to the medium, according to a probability distribution $D(\varepsilon)$ known as the jet energy-loss distribution or a \textit{quenching weight} (QW). This leads to a shift in the $\pT$ spectrum, i.e. $\pT + \varepsilon \mapsto \pT$. The energy-loss distribution depends both on medium and on jet properties,  i.e. $D(\varepsilon) = D\big(\varepsilon| \{\hat q, T, L,\ldots \};\, \{\pT, R,\ldots \} \big)$, which fluctuate from event to event. Theoretical models have been proposed to describe $D(\varepsilon)$, see e.g. \cite{Salgado:2003gb,Arleo:2002kh}, and to use Bayesian inference to model it \cite{He:2018gks, Zhang:2022rby, Zhang:2023oid, Xing:2023ciw, Wu:2023azi}.

Bypassing event-by-event fluctuations, we will treat the spectrum in Eq.~\eqref{eq:fact}, and the resulting energy-loss distribution $D(\varepsilon)$, as an event-averaged quantity. We focus currently only on one specific centrality of Pb-Pb collisions (i.e. 0--10\% centrality), assuming that the fluctuating medium properties from event to event are captured by this universal description when considering event-averaged observables. In order to keep a minimal parametrization, we also neglect possible dependencies on jet scales, in particular on $\pT$, see also Sec.~\ref{sec:theory-poisson}, and the data we analyze is all from reconstructed jets with a fixed cone parameter $R$. However, a crucial ingredient in our setup is to treat the quenching of quark- and gluon-initiated jets independently. We wish to test whether this ``minimal scenario'' provides a consistent description of the experimental data. Our procedure is thenceforth fully data-driven and provides a complementary view on the role of bias effects for hard observables in heavy-ion collisions.

It is important to stress that instead of inferring a range of parameter values for any particular model, we aim to check consistency among existing measurements of jet suppression and to extract generic features. Our first task is therefore to robustly test the hypothesis in Eq.~\eqref{eq:fact}. This is something we refer to as \textit{universality} of the QW, in the same spirit as, e.g., the universality of the soft function in jet measurements. While this property is widely used in phenomenological models, it has not been scrutinized in a data-driven fashion before. Our second task is to study the color-dependence of jet suppression. To achieve this aim, we have considered subsets of existing data to check the constraining power it has for both quark- and gluon-jet quenching. Our findings imply that bias effects complicate this task considerably, and that either \texttt{i)} experimental data less affected by bias has to be included in the analysis, or \texttt{ii)} further assumptions on the prior have to be made to reach a clear conclusion. Pursuing the second option, we define the color ration $C_R$ as the ratio of mean energy losses of gluon- and quark-initiated jets, i.e.
\begin{equation}
    \label{eq:color-ratio}
    C_R = \frac{\langle \varepsilon_g \rangle_\text{jet}}{\langle \varepsilon_q \rangle_\text{jet}} \,.
\end{equation}
Based on the analysis of experimental data we find evidence of super-Casimir scaling $C_R > N_c/C_F$, see Sec.~\ref{sec:prior-constraint}. This indicates that gluon-jets are more sensitive to quenching than quark-jets, possibly due to non-linear effects of jet energy loss due to secondary jet emissions.

The paper is structured as follows. In Sec.~\ref{sec:theory} we introduce the main notions of QCD processes and discuss the expected features of $D(\varepsilon)$ for a stochastic, Poisson-like description of jet quenching. We also discuss the three parametrizations of $D(\varepsilon)$ that we employ in our data analysis. In Sec.~\ref{sec:data}, we discuss the two classes of experimental data we have selected for our task, namely the inclusive jet spectrum and photon-jet coincidence spectra. Next, we discuss the setup for the Bayesian inference in Sec.~\ref{sec:bayesian}, including a discussion of closure tests, before moving on to the inference on experimental data in Sec.~\ref{sec:inference}. Our main results on the universality of jet quenching can be found in Secs.~\ref{sec:universality}. Realizing that the extraction of color dependence is highly sensitive to biases we have also suggested a more restricted search within the priors, see Secs.~\ref{sec:insight} and \ref{sec:prior-constraint}, respectively. Finally, we summarize our findings and provide an outlook in Sec.~\ref{sec:conclusions}.

\section{Jet energy loss in the quark-gluon plasma}
\label{sec:theory}

Partons traversing a hot and dense nuclear medium are affected by elastic and inelastic scattering. For a sufficiently large medium, multiple emissions can take place occur and the radiated quanta can cascade via secondary elastic and inelastic interactions. In course of these processes, a significant energy fraction of the original parton is transferred to particles with energies around the temperature scale of the medium, i.e. $\omega \sim Q_\med$ with $Q_\med \sim T$, and moving at large angles compared to the original direction. Some of the lost energy thermalize with the surrounding medium and can dissipate back into the jet cone \cite{Casalderrey-Solana:2006lmc, Qin:2009uh, Tachibana:2017syd, Chen:2017zte}. The amount of energy ending up outside the cone of the reconstructed jet is referred to as the energy-loss of the same jet. Since for high-$\pT$ jets the elastic energy loss is much smaller than the inelastic energy loss \cite{Qin:2007rn,Cao:2016gvr}, we will focus on the latter for the purposes of the discussion in this Section.

However, the hard QCD processes producing energetic particles simultaneously open the phase for emissions due to the collinear and soft divergences of QCD \cite{Mehtar-Tani:2017web,Caucal:2018dla}. For final-state particles, the QCD evolution \cite{Dasgupta:2014yra, Dasgupta:2016bnd, Kang:2016mcy} brings the processes at the scale of the hard cross-section $\sim \pT$ down to the scale of the reconstructed jet $Q_\text{\tiny jet} \sim \pT R$, where $R$ is the jet cone-angle. Subsequent splittings inside the jet span a wide range of characteristic formation times, from the initial time-scales of the order $\sim 1/(\pT R)$ to the hadronization time-scale $\sim \pT R/Q_0^2$, where $Q_0 \sim \Lambda_\text{\tiny QCD}$ is the hadronization scale.

If the jet scale is large compared to medium scales, i.e. $\pT R \gg Q_\med$, or naively $1/(\pT R) \ll L$ indicating that the emission takes place inside the medium, collinear emissions can take place on time-scales much shorter than the ones typically involved in medium-induced processes. The initially collinear vacuum-like splittings are resolved by the medium when their transverse extent is resolved by multiple-scattering in the medium \cite{Mehtar-Tani:2017web}. The system that starts interacting with the medium is therefore rarely a single parton but rather a partially developed shower consisting of a collimated set of partons. This enhances the effects of energy-loss since each resolved parton acts as an incoherent contributor to the total energy-loss at large angles. 

For inclusive observables, that only depends on the $\pT$ and $R$ of the measured jets, the information about medium interactions enter only as the amount of lost energy $\varepsilon$ fluctuates on an jet by jet level. Hence, the resulting energy loss distribution $D(\varepsilon)$ depends on a wide range of factors, such as on the initial jet QCD evolution, on the interaction of the jet with the medium, and on the properties of the medium itself, such as its length and temperature. 

Our two working hypotheses are therefore:
\begin{enumerate}
    \item The energy loss distribution is an approximately universal quantity for a range of jet $\pT$, keeping the cone angle $R$ fixed, and for similar medium profiles (kept approximately constant by fixing the centrality class).
    \item Jet energy is dominated by Poissonian processes, i.e. independent, stochastic processes which add up to the total energy loss.
\end{enumerate}
We will test the first hypothesis against existing experimental data, examining whether the evidence supports or contradicts it. The second hypothesis is instrumental to interpret the color dependence of jet quenching, which is the aim of the second part of the paper.

\subsection{Poisson-like distributions}
\label{sec:theory-poisson}

To understand these dependencies, we can start by taking a look at the case where a jet is reduced to a single parton which, under certain assumptions, has a simple theoretical description. Neglecting for the moment elastic contributions, the interaction of a hard parton with the medium induces gluon bremsstrahlung  which is responsible for decreasing the energy of the leading parton. This is described by the spectrum $\rmd I/\rmd \omega$, where $\omega$ is the energy of the emitted gluon. We can then construct a probability distribution for a single parton to loose energy $\varepsilon$, which we will denote $D^{(1)}(\varepsilon)\equiv\mathcal{D}(\varepsilon)$ to distinguish it from the energy-loss distribution of a full \textit{jet}, see below. Assuming that the bremsstrahlung gluons are emitted independently, the all-order energy loss distribution is a sum of Poisson-like contributions,
\begin{equation}
    \label{eq:single-parton-energy-loss}
     \mathcal{D}(\varepsilon) = \rme^{-\int_0^\infty\!\!\dd\omega\, \frac{\dd I}{\dd\omega}}
    \sum_{n=0}^\infty\frac{1}{n!}\, \prod_{i=1}^n\int_0^\infty\dd\omega_i \, \frac{\dd I}{\dd\omega_i} \,
    \delta\!\left(\varepsilon-\sum_{i=1}^n\omega_i\right) \,,
\end{equation}
see also \cite{Baier:2001yt,Salgado:2003gb}.
In this distribution, the parameter $\varepsilon$ is continuous but it has to be the sum of a discrete number of individual emissions $\omega_i$.
This distribution is normalized $\int_0^\infty \rmd \varepsilon D(\varepsilon) = 1$, and its moments are directly linked to underlying bremsstrahlung distribution, e.g., 
\begin{align}
    \label{eq:average-eps}
    \langle \varepsilon \rangle_1 &= \langle \omega \rangle  \,,\\
    \langle \varepsilon^2 \rangle_1 &= \langle \omega^2 \rangle + \langle \omega \rangle^2 \,, 
\end{align}
where $\langle \omega^n \rangle = \int_0^\infty \rmd \omega \, \omega^n \rmd I/\rmd \omega$. The subscript denotes that we are referring to the single-parton energy loss distribution. Note that the existence of higher moments is here guaranteed by the restriction that $\omega \leq E$, however this is not necessarily the case for a general, normalized probability distribution \footnote{These results generalize straightforwardly when including elastic energy losses as $\mathcal{D}(\varepsilon) = {\protect \int} \rmd \varepsilon_1 {\protect \int} \rmd \varepsilon_2 \, \delta(\varepsilon- \varepsilon_1 - \varepsilon_2) \mathcal{D}_\text{rad}(\varepsilon_1) \mathcal{D}_\text{el}(\varepsilon_2)$, such that $\langle \varepsilon\rangle = \langle \varepsilon_\text{rad} \rangle_1 + \langle \varepsilon_\text{el} \rangle_1$, and so on. Here $\mathcal{D}_\text{rad}(\varepsilon_1)$ and $\mathcal{D}_\text{rad}(\varepsilon_1)$ are the energy loss distributions due to radiative and elastic processes, respectively.}.

Finally, a comment on the flavor-dependence of energy loss is in place. The basic building block of the energy-loss distribution, namely the radiative spectrum $\rmd I/\rmd \omega$, has to be proportional to the QCD coupling constant $\alpha_s$ and the color factor of the emitting particle, i.e. $\rmd I/\rmd \omega \propto \alpha_s C_i$, where $C_i = C_F$ for quarks ($i=q$) and $C_i = N_c$ for gluons ($i=g$). Then, the distribution defined in Eq.~\eqref{eq:single-parton-energy-loss} can straightforwardly be assigned to a parton with flavor $i$ by $\mathcal{D}(\varepsilon) \mapsto \mathcal{D}_i(\varepsilon)$. It follows that the mean energy energy loss exhibits the same scaling, namely $\langle \varepsilon_i \rangle_1 \propto C_i$.

To focus for a moment on a concrete example of a radiative spectrum, let us consider a simple model where the spectrum is dominated by multiple-scattering \cite{Baier:1996kr}, i.e. $\rmd I/\rmd \omega = \bar \alpha \sqrt{\hat qL^2/\omega^3}$ for $T < \omega$, where $T$ is the medium temperature, $L$ is the distance traversed in the QGP and $\bar \alpha = \alpha_s C_i/\pi$. Here, $\hat q \propto T^3$ is the jet quenching parameter. This is a good model for $E < \hat q L^2$, but at $E > \hat q L^2$ we expect a more rapid drop of the spectrum (due to single hard scattering processes in the medium). We can therefore restrict the upper range of the emitted energy as $\omega < \min(E,\hat qL^2)$. Considering a large separation of jet and medium scales, $E \gg \hat qL^2$, we get $\langle\omega\rangle=2\bar\alpha\hat q L^2$. With minimal approximations, the energy loss distribution turns out to be $\mathcal{D}^\text{\tiny LPM}(\varepsilon) \simeq \sqrt{\omega_s/\varepsilon^3}\rme^{- \pi\omega_s/\varepsilon}$, where $\omega_s = \bar \alpha^2 \hat q L^2$. The maximum of this distribution is readily found to be at $\varepsilon_\text{max} = \frac23 \pi \omega_s$. Using the relation in \eqref{eq:average-eps}, we therefore find 
\begin{equation}
    \label{eq:mean_mode_theory}
    \frac{\langle \varepsilon \rangle}{\varepsilon_{\rm max}} \sim \frac1{\alpha_s} \,.
\end{equation}
This scaling should also approximately hold on general grounds for a wider class of Poisson-like models for energy loss. This scaling behavior will play an important role in constraining the parameter space for Bayesian inference, described in Sec.~\ref{sec:prior-constraint}, especially for fat-tailed distributions, where the first moment can be significantly displaced compared to the maximum of the distribution.

Assuming a strong separation between jet and medium scales, at leading-logarithmic approximation \cite{Mehtar-Tani:2017web,Caucal:2018dla}, see also \cite{Abreu:2024wka}, leads therefore to a sharply divided two-stage picture where the jet initially fragments into a set of sub-jets which, subsequently, are resolved by the medium and can further contribute to energy loss. Let us then consider a jet with $n$ partons resolved by the medium. Assuming small energy losses compared to the total jet $\pT$, it is justified to consider independent energy loss off all partons. Such an object consists, at leading logarithmic accuracy, of the parent parton, with flavor $i = q,g$, and a set of radiated soft gluons. This results in the $n$-parton energy loss distribution for a jet with flavor $i$,
\begin{multline}
    D_i^{(n)}(\varepsilon) = \int_0^\infty \rmd \varepsilon_1\ldots \int_0^\infty \rmd \varepsilon_n\, \mathcal{D}_i(\varepsilon_1) \\ \times \mathcal{D}_g(\varepsilon_2) \ldots \mathcal{D}_g(\varepsilon_{n}) \delta\left(\varepsilon- \sum_{i=1}^n \varepsilon_i\right) \,.
\end{multline}
This implies that the average energy loss of a jet consisting of a leading parton of flavor $i$ and $n_i$ gluons, $\langle \varepsilon_i \rangle_\text{jet} \simeq \langle \varepsilon_i \rangle_{1+n_i}$, scales as
\begin{equation}
    \label{eq:color-scaling-0}
    \langle \varepsilon_i \rangle_{1+n_i} = \langle \varepsilon_i \rangle \, \left(1 + n_i \frac{N_c}{C_i} \right)\,.
\end{equation}
In this discussion we have placed a flavor index and therefore a comment on the size of $n_i$ is in place. Qualitatively, since $n_i$ counts the number of vacuum-like emissions inside the medium, we expect it to scale as $n_i \sim \alpha_s C_i \int \rmd \Pi_\text{in}$, where $\int\rmd\Pi_\text{in}$ is the available phase space \cite{Mehtar-Tani:2017web, Mehtar-Tani:2021fud}. Considering only the double-logarithmic limit of splitting functions in vacuum, and imposing some condition on the emission phase space, would lead to $\int \rmd \Pi_\text{in} \sim \ln \pT$. Hence, although the number of such emissions is small due to the smallness of $\alpha_s$, it should be partly compensated by the phase space. Realistic calculations \cite{Mehtar-Tani:2021fud,Mehtar-Tani:2024jtd} indicate that $n_i \lesssim 2-3$. These considerations only account for the \textit{primary} gluon emissions off the leading parent parton. \textit{Secondary} emissions can also take place, leading to a fully non-linear dependence of jet energy loss. Since gluon jets radiate more, these non-linear effects are more pronounced.

Finally, we would like to investigate the color dependence of jet energy loss through the color factor $C_R$, defined in Eq.~\eqref{eq:color-ratio}. Since the number of \textit{primary} vacuum-like gluons, or equivalently the phase space for in-medium radiation, scales with the color factor then $n_i/C_i$ in \eqref{eq:color-scaling-0} is the same for both quark- and gluon-initiated jets.
Hence, we find that
\begin{align}
    \label{eq:Casimir}
    C_R> \frac{\langle \varepsilon_g \rangle_{1+n_g}}{\langle \varepsilon_q\rangle_{1+n_q}} = \frac{\langle \varepsilon_g \rangle_{1}}{\langle \varepsilon_q\rangle_{1}} = \frac{N_c}{C_F} = 2.25\,.
\end{align}
Since both vacuum-like and medium-induced emissions scale with the relevant color factor, the baseline prediction is trivial. Note that our discussion only accounted for \textit{primary} emissions so the estimate sets Casimir scaling $N_c/C_F$ as a lower-bound for the color ratio $C_R$ of full jets, under our working hypotheses of independent energy loss.

\subsection{Data-driven approach}
\label{sec:parametrizations}

Having discussed some characteristic features for the Poissonian-like model of radiative energy loss, let us now turn to the data-driven approach pursued in the remainder of this paper. The full energy loss distribution for jets, as defined in Eq.~\eqref{eq:fact}, only assumes a separation of jet and medium scales which implies that the total lost energy $\varepsilon$ is smaller than the jet scale $\pT$. No further assumption is being made on the mechanism leading to the lost energy. In this work we also limit ourselves to considering the most central heavy-ion collisions, which fixes the average geometry of the problem.

In order to investigate a wide range of parametric behaviors, we will use three examples from a 2-parameter familty of analytical distributions as parametrizations of the jet energy loss distribution $D(\varepsilon)$. The three are discussed below.
\begin{description}
\item[Normal parametrization]
Uses the normal distribution, with the standard parameters representing the ``mean'' value $\mu$ and standard deviation $\sigma$, reads
\begin{equation}
    D(\varepsilon| \mu, \sigma) = \frac{\mathcal{N}}{\sigma \sqrt{2\pi}} \rme^{-\frac{(\varepsilon-\mu)^2}{2\sigma^2}} \,,
\end{equation}
where the additional normalization factor, $\mathcal{N}^{-1}=\big[1+\text{erf}\big(\mu/(\sqrt{2}\sigma)\big)\big]/2$, is necessary to ensure that the distribution is normalized over $\varepsilon \in (0,\infty)$. The average of the distribution,
\begin{equation}
    \label{eq:mean_normal}
    \langle \varepsilon \rangle = \mu + \sqrt{\frac{2}{\pi}}\frac{\sigma\, \rme^{-\mu^2/(2\sigma^2)}}{1+\text{erf}\big(\mu/(\sqrt{2}\sigma)\big)} \,,
\end{equation}
now differs slightly from the \textit{maximum} of the distribution, located at $\varepsilon_\text{max} =\mu$.

\item[Log-normal parametrization]
Uses the log-normal distribution, which reads
\begin{equation}
    D(\varepsilon|\mu,\sigma) = \frac{1}{\varepsilon \sigma \sqrt{2\pi}}\rme^{-\frac{(\ln \varepsilon - \mu)^2}{2 \sigma^2}} \,.
\end{equation}
In this case, the average energy loss is
\begin{equation}
    \label{eq:mean_lognormal}
    \langle \varepsilon \rangle = \rme^{\mu+ \sigma^2/2}\,,
\end{equation}
while the \textit{maximum} of the distribution is located at $\varepsilon_\text{max} = \rme^{\mu - \sigma^2}$. Also, note that
\begin{equation}
    \label{eq:mean_max_lognormal}
    \frac{\langle \varepsilon \rangle}{\varepsilon_{\rm max}} = \rme^{\frac32 \sigma^2} \,.
\end{equation}
The exponential implies a strong sensitivity on the parameter $\sigma$. This indicates a strong effect from the behavior of the parametrization in the tail (so called ``fat tail'').

\item[Gamma parametrization]
The final case we include, is when the energy loss distribution is parametrized as a gamma distribution, and reads
\begin{equation}
    D(\varepsilon|\alpha, \theta)=\frac{\theta^{-\alpha}\varepsilon^{\alpha-1}e^{-\varepsilon/\theta}}{\Gamma(\alpha)} \,,
    \label{eq:gamma_dist}
\end{equation}
which is parametrized in terms of the shape parameter $\alpha$, and the scale $\theta$. Due to its nice convolution properties, in the Poisson approximation for soft gluon emissions discussed above, $\alpha$ is directly related with the number of emitted gluons $\alpha=1+ n$ and the scale can be identified as the mean energy that each emitted gluon takes from the jet cone to the medium, $\theta \sim \langle \omega \rangle$.

The gamma distribution has also the asset of having a well defined and simple mean, given by
\begin{equation}
    \label{eq:mean_gamma}
    \langle\varepsilon\rangle=\alpha\theta\,,
\end{equation}
as expected. The change of variables $\varepsilon\mapsto x\equiv\varepsilon/\langle\varepsilon\rangle$ allows us to write the gamma distribution directly in terms of $\langle\varepsilon\rangle$,
\begin{equation}
    D(x|\langle\varepsilon\rangle,\alpha)=\frac{\alpha^{\alpha}x^{\alpha-1}e^{-\alpha x}}{\Gamma(\alpha)}\,.
    \label{eq:gamma_dist}
\end{equation}
We also immediately find the maximum distribution, resulting in 
\begin{equation}
    \frac{\langle \varepsilon \rangle}{\varepsilon_\text{max}} = \frac{\alpha}{\alpha-1} \,.
\end{equation}

\end{description}

For each of these three parametrizations, we need an instance applied to quark-initiated jets $D_q(\varepsilon)$ and to gluon-initiated jets $D_g(\varepsilon)$ extending the number of parameters to 4. We summarized the information about the parametrizations in Tab.~\ref{tab:parametrizations}. For the main part of the paper, no prior information is assigned to the parameters and they are assigned a flat prior in a wide range. In Sec.~\ref{sec:prior-constraint}, however, we restrict the prior range to enforce the physical insight based on Eq.~\eqref{eq:mean_mode_theory}. Other energy loss parametrizations have been suggested, including jet $\pT$ dependence \cite{He:2018gks, Zhang:2023oid, Wu:2023azi} and additional path-length dependence \cite{Kumar:2024yzj,Spousta:2016agr,Ogrodnik:2024qug}. To keep our approach as model-agnostic and economic in terms of introduced parameters, we have not considered implementing these additional dependencies in our parametrizations.

The Bayesian inference on experimental data aims to determine the valid ranges for the parameters of these distributions. Since these parameters represent the average effect of numerous fluctuations and confounding factors for a given jet, as described above, we should avoid assigning their central values a direct physical interpretation. Instead, our goal is to look for consistencies across data sets that may reveal generic scaling behavior.

\begin{table}
    \renewcommand{\arraystretch}{1.3}
    \centering
    \begin{tabular}{c|c}
         parametrization & inferred parameters $\theta$  \\
         \hline
         normal & $\mu_q$, $\mu_g$, $\sigma_q$, $\sigma_g$ \\
         log-normal & $\mu_q$, $\mu_g$, $\sigma_q$, $\sigma_g$ \\
         gamma & $\langle \varepsilon_q\rangle$, $\langle \varepsilon_g \rangle$, $\alpha_q$, $\alpha_g$
    \end{tabular}
    \caption{Summary of parametrizations and their parameters used for inference.}
    \label{tab:parametrizations}
\end{table}

\section{Observables and kinematic processes}
\label{sec:data}

\begin{figure}[t]
    \centering
    \includegraphics[width=0.9\columnwidth]{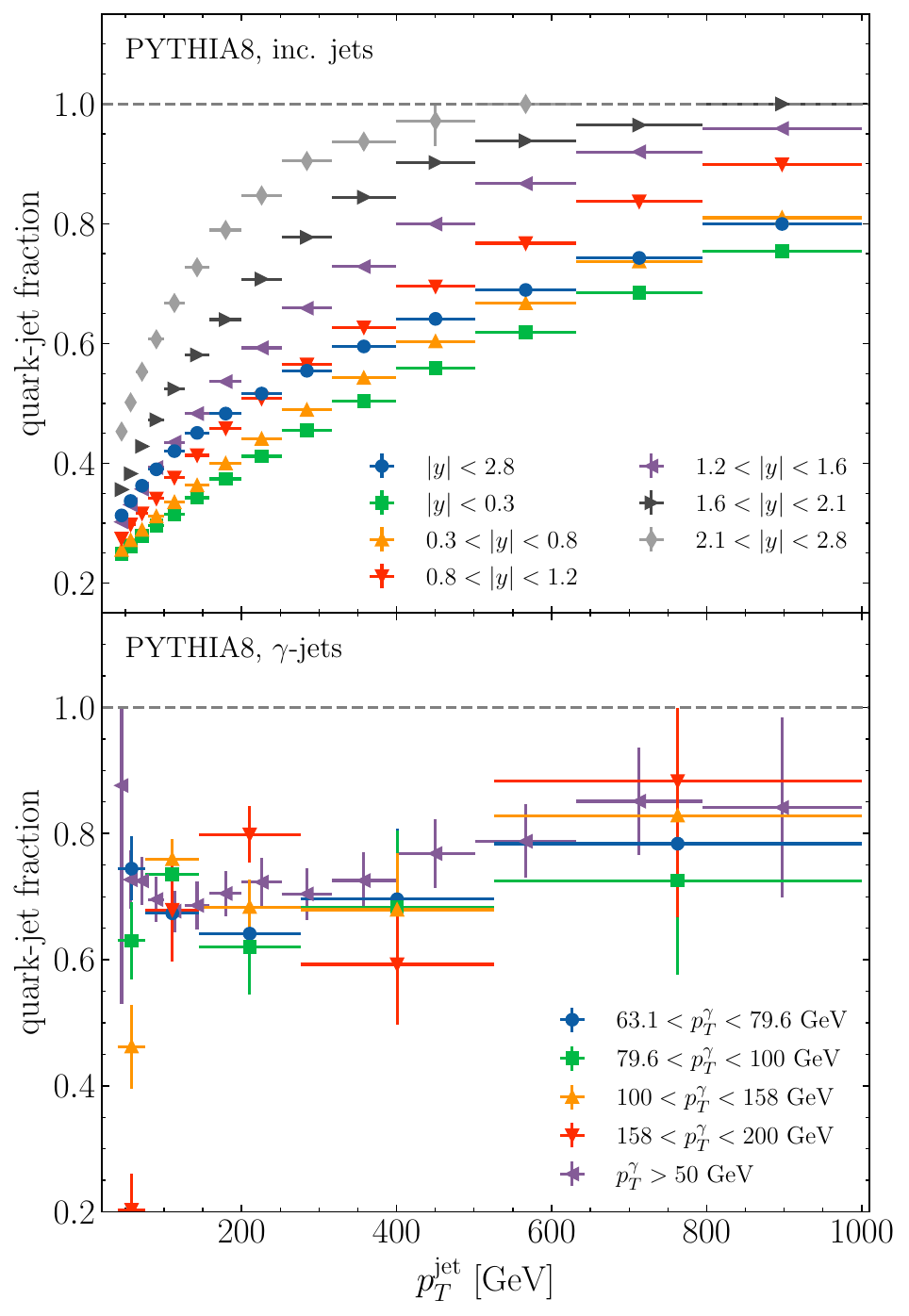}
    \caption{Upper panel: fraction of inclusive jets initiated by quarks as a function of the jet $p_T$ for the rapidity $|y|$ intervals of interest, from PYTHIA8 samples. Lower panel: fraction of photon-tagged jets initiated by quarks as a function of the jet $p_T$ for the five $p_T^\gamma$ intervals of interest, also from PYTHIA8 samples.}
    \label{fig:qrk_frac}
\end{figure}

To address the problems set out in the Introduction, we perform a data-driven Bayesian inference on experimental data.
Observables are chosen for two classes of jet events: inclusive jet events ($pp/AA\rightarrow\text{jet}+X$), and photon-tagged jet events ($pp/AA\rightarrow\gamma+\text{jet}+X$). As alluded to above, the choice of these two particular observables are motivated by two factors, namely:
\begin{itemize}
    \item the observables have different sensitivity to bias effects. The jet spectrum is steeply falling as a power law in $\pT$, while the $\xjg$ distribution measured balanced jet-photon configurations at $\xjg \approx 1$.
    \item the observables probe different fractions of quark- and gluon-initiated jets. 
\end{itemize}
The inclusive jet spectrum is dominated by gluon-initiated jets at $\pT \lesssim 150$ GeV, with quark-initiated jets only taking over at $\pT > 200-300$ GeV. A division into different rapidity intervals give access to additional quark-/gluon-jet fractions, with gluon-initiated jets dominating at mid-rapidity, and quark-initiated jets taking over at decreasing momentum as we go forward in rapidity. The inclusion of rapidity dependence in this context is important to clarify the distinction between the energy lost by quark- and gluon-initiated jets, see also \cite{Pablos:2022mrx}. Photon-tagged jet processes are naturally dominated by quark-initiated jets from Compton scattering. Figure~\ref{fig:qrk_frac} shows the fraction of quark-initiated jets as a function of the jet $\pT$, here denoted $\pTj$ , obtained for the two classes of jet observables in study.

In this work, we look at the following jet observables:
\begin{itemize}
    \item the jet-inclusive nuclear modification factor $R_{AA}$ as a function of $\pT$, and the correspondent relative $R_{AA}$ for different rapidity bins, $R_{AA}^y/R_{AA}^{|y|<0.3}$; and
    \item the per-photon yield of photon-tagged jets, $1/N_\gamma\cdot\dd N^{AA}/\dd\xjg$, for different $p_T^\gamma$ bins.
    \item the photon-tagged jet nuclear modification factor $R_{AA}$ for $\pTg > 50$ GeV at mid-rapidity.
\end{itemize}
The jets are reconstructed with the anti-$k_t$ algorithm with cone-parameter $R=0.4$. The measurements, done by the ATLAS collaboration, can be found in \cite{ATLAS:2018gwx,ATLAS:2018dgb,ATLAS:2023iad}, for for Pb-Pb central collisions (centrality 0-10\%), and $\sqrt{s}=5.02$ TeV. Only one centrality and center of mass energy is chosen across the observables, in order to ensure that the QGP in the different events presents as close properties as possible, so the average of $D(\varepsilon)$ over the medium properties is valid. Detector effects are not considered, since data analysis from ATLAS includes unfolding.

\subsection{Inclusive jet observables}
\label{sec:inclusive_observables}

The nuclear modification factor $R_{AA}$ is defined as
\begin{equation}
    \label{eq:raa}
    R_{AA}(\pT)=\frac{\sigma^{AA}(\pT)}{\sigma^{pp}(\pT)}\,.
\end{equation}
Being an observable of fundamental interest, we will give a brief account of how we compute the inclusive jet spectrum, denoted as $\sigma^{pp}(\pT,R) = \rmd \sigma^{pp}/(\rmd \eta \rmd \pT)$ for a specific binning in $\pT$ and pseudorapidity $\eta$ and for a cone resolution parameter (cone-angle) $R$. The spectrum can be written as \cite{Dasgupta:2014yra}
\begin{equation}
    \sigma^{pp}(\pT,R) = \sum_{k=q,g} f_{\text{jet}/k}^{(n_k-1)}(R| \pT,R_0) \hat \sigma_k(\pT,R_0) \,,
\end{equation}
where $R_0$ is the angular scale related to the hard collision (in practice, $R_0 \sim 1$ such that $Q_\text{hard} \sim \pT$). Here, $n_k\equiv n_k(\pT,R_0)$ is the power-index of the cross-section of the initial hard parton $k$, namely $\hat \sigma_k(\pT,R_0)$. In collinear factorization, this object is given as $\hat \sigma_k = \sum_{i,j,l}\, f_{i/p}\otimes f_{j/p} \otimes \hat \sigma_{ij \to kl}$, where the partonic, hard cross-section $\hat \sigma_{ij \to kl}$ is summed over all unmeasured variables. Finally, $f^{(n)}_{\text{jet}/k}(R|\pT,R_0)$ is the moment of the semi-inclusive jet fragmentation function of an initial hard parton $k$. We implement the effects of energy loss due to medium interactions as
\begin{equation}
    \label{eq:sigma-AA}
    \sigma^{AA}(\pT,R) = \sum_{i=q,g} \int_0^\infty \rmd \varepsilon \, D_i(\varepsilon) \tilde \sigma^{pp}_i(\pT+\varepsilon,R) \,,
\end{equation}
where the $\tilde\sigma^{pp}$ is the baseline cross-section with the replacement of the proton PDFs by nuclear PDFs (nPDFs). From Eq.~\eqref{eq:raa}, the nuclear modification factor then becomes
\begin{equation}
    R_{AA}(\pT,R) = \frac{1}{\sigma^{pp}(\pT,R)}\sum_{i=q,g} \int_0^\infty \rmd \varepsilon \, D_i(\varepsilon) \tilde \sigma^{pp}_i(\pT+\varepsilon,R)\,.
\end{equation}
For more details, see \cite{Dasgupta:2014yra} and \cite{Mehtar-Tani:2021fud,Mehtar-Tani:2024jtd}.

\begin{figure*}
    \centering
    \includegraphics[width=\textwidth]{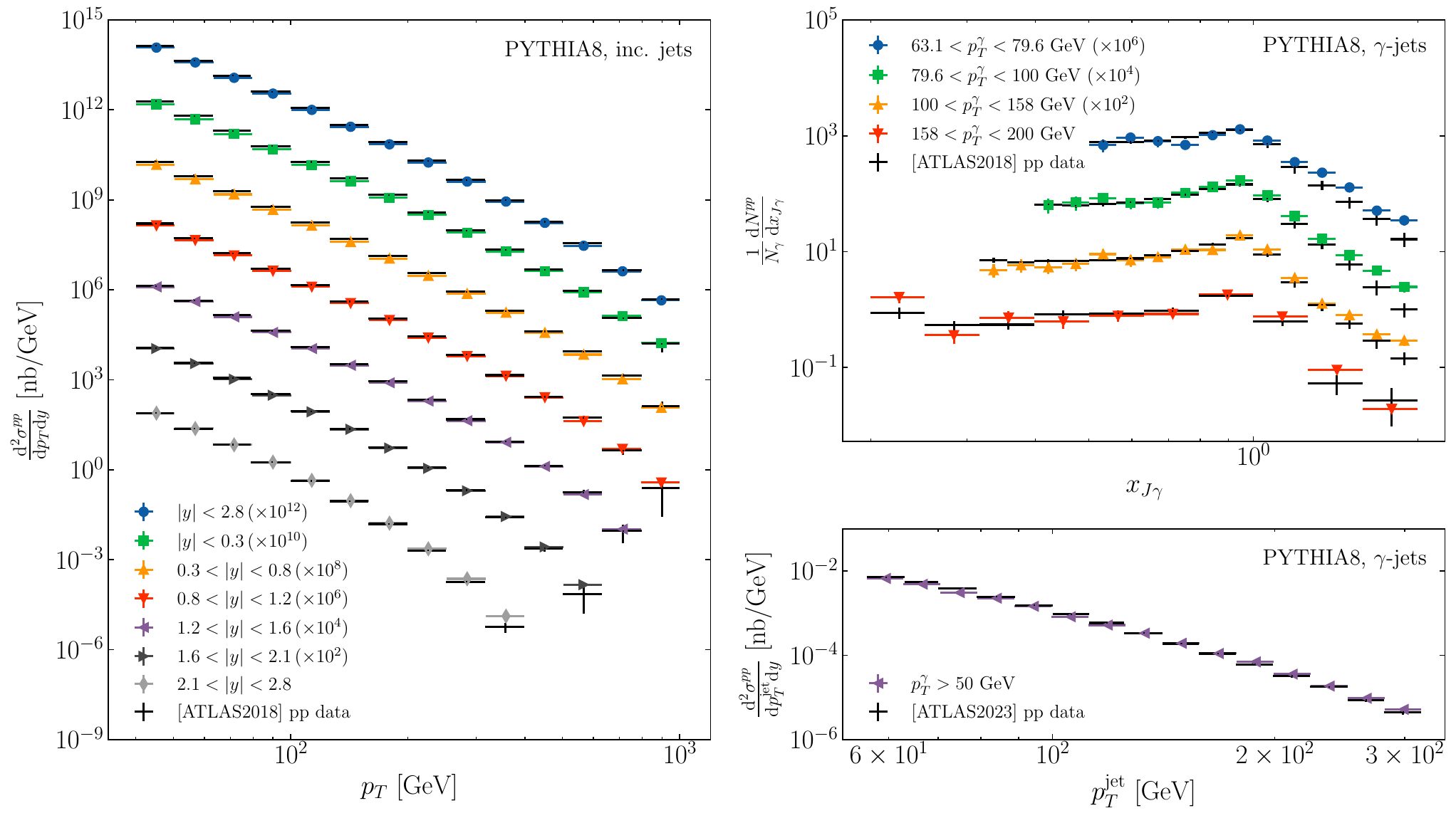}
    \caption{Left panel: inclusive jet cross-section as a function of jet $p_T$ for the rapidity $|y|$ intervals of interest, in $pp$ collisions sampled using PYTHIA8. Left panels: photon-tagged jet yield (upper panel) as a function of jet imbalance $\xjg$, and cross-section (lower panel) as a function of jet $p_T$ for the $p_T^\gamma$ intervals of interest, in $pp$ collisions sampled using PYTHIA8. All points are compared with the corresponding data points measured by the ATLAS experiment.}
    \label{fig:pythia_pp}
\end{figure*}

\begin{figure*}
    \centering
    \includegraphics[width=\textwidth]{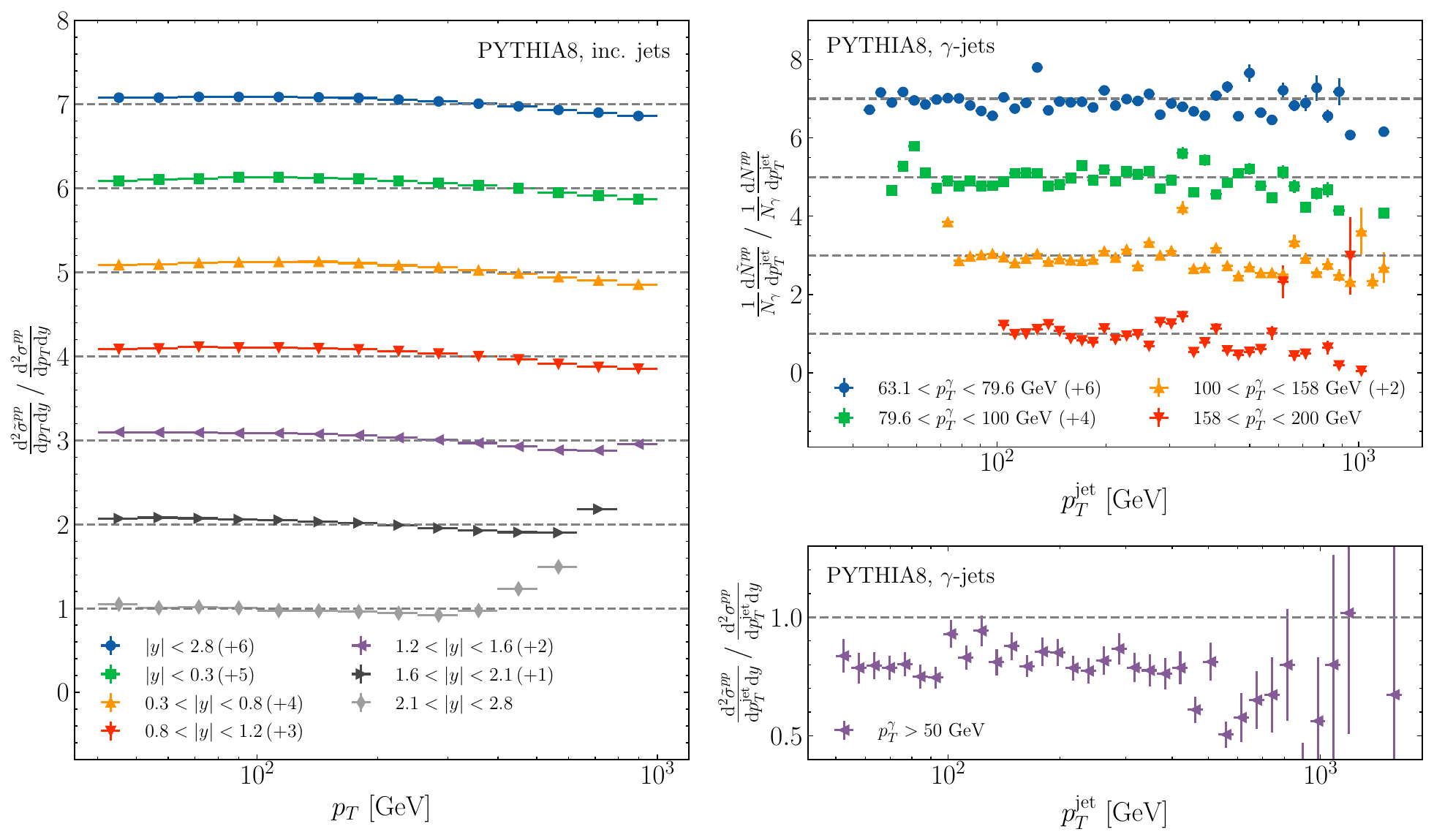}
    \caption{nPDF effect on the inclusive jet cross-section (left panel), and on the photon-tagged jet yield (upper left panel) and cross-section (lower left panel), as functions of $p_T$. Points obtained from PYTHIA8 samples.}
    \label{fig:nPDF_effect}
\end{figure*}

In practice, the $pp$ spectra $\sigma^{pp}$ and $\tilde\sigma^{pp}$ are obtained via Monte Carlo event generation. Jets are generated using the leading-order event generator PYTHIA8 \cite{Sjostrand:2006za,Sjostrand:2007gs} and reconstructed with the anti-$k_t$ algorithm with $R=0.4$ using FastJet \cite{Cacciari:2011ma}. $\tilde\sigma^{pp}$ is obtained by including the nPDF set EPS09 \cite{Eskola:2009uj} at the generator initialization. The generated $pp$ spectrum $\sigma^{pp}$ is shown in Fig.~\ref{fig:pythia_pp}, for the rapidity cuts of interest, together with the measured spectrum from ATLAS \cite{ATLAS:2018gwx}. The agreement between the simulation and the data guaranties a correct  baseline. Due to the small difference between the generated spectra with and without nPDFs, the effect of the nPDFs on the $pp$ spectrum is shown in Fig.~\ref{fig:nPDF_effect}, by plotting the ratio $\tilde\sigma^{pp}(\pT)/\sigma^{pp}(\pT)$. The quark- and gluon-initiated jet spectra, $\tilde\sigma_q$ and $\tilde\sigma_g$ respectively, are obtained by matching the jet to outgoing partons from the LO hard matrix element using several, internally consistent methods. The fraction of quark-initiated jets in $pp$ collisions, depicted in Fig.~\ref{fig:qrk_frac}, is generated for the kinematic cuts and tags relevant for the observables in study. We see that the fraction varies strongly both with $\pT$ and rapidity. While the most precise data on jet suppression is currently at mid-rapidity around $\pT \sim 150-200$ GeV, a combined analysis of several kinematic regimes should be sufficient to take advantage of these variations to properly separately constrain quark- and gluon-initiated jet energy loss.

The histograms are fitted by continuous functions to facilitate numerical integration later on, for which high integration limits demand a precise extrapolation of the spectra to high-$\pT$. The inclusive jet spectrum is known to steeply fall with $\pT$ \cite{Ellis:1996mzs}, hence following 
\begin{equation}
    \label{eq:fall_spec}
    \sigma_i^{pp}(\pT,R)= A^{(i)}_0\left(\frac{\pT^0}{\pT}\right)^{n_i(\pT)}\,,
\end{equation}
where $A_0^{(i)}$ and $\pT^0$ are constants, and the flavor-dependent spectral index $n_i(\pT)$ is approximated as a power-series in the logarithm of $\pT$, i.e. $n_i(\pT) = \sum_{j=0}^N n_i^{(j)} \ln^j(\pT/\pT^0)$, see, e.g., \cite{Spousta:2015fca} . The free parameter $\pT^0$ is set as the lowest $\pT$ considered, and $A_0^{(i)}$ and $n_i^{(j)}$ (truncated at $N=5$) are fitted for quark- ($i=q$) and gluon-initiated ($i=g$) jet spectra for a given $R$. The resulting curves for both $\sigma^{pp}$ and $\tilde\sigma^{pp}$ are shown in the figures of App.~\ref{sec:spectra_fits}. The same fitting procedure is performed on the generated quark- and gluon-initiated jet spectra, where Eq.~\eqref{eq:fall_spec} was shown to also apply to these spectra.

Finally, it is worth commenting on the characteristic feature of convolutions such as in Eq.~\eqref{eq:sigma-AA}. Due to the steeply falling behavior of the spectrum in $\pT$, the spectrum in heavy-ion collisions can be approximated by
\begin{equation}
    \sigma^{AA} (\pT,R) \simeq \sum_{i=q,g} \tilde\sigma_i^{pp}(\pT,R) \tilde D_i(n/\pT) \,,
\end{equation}
where $\tilde D_i(\nu)$ is the Laplace transform of the energy-loss distribution,
\begin{equation}
    \tilde D(\nu) = \int_0^\infty \rmd \varepsilon\, D(\varepsilon) \rme^{-\nu \varepsilon} \,,
\end{equation}
with Laplace variable $\nu = n/\pT$ \footnote{Here we have used the standard approximation, i.e. ${\protect \int}_0^\infty \varepsilon \, D(\varepsilon)/(\pT+\varepsilon)^n \simeq \pT^{-n} {\protect \int}_0^\infty \rmd \varepsilon \, D(\varepsilon)\rme^{-n \varepsilon/\pT}$.}. This form highlights how the survival bias has a strong effect on the extracted energy loss distribution. For $\varepsilon > \nu^{-1}$ the contributions to $\tilde D(\nu)$, or, in other words, the $R_{AA}$, are exponentially suppressed. This indicates an inherent lack of information about large energy loss processes in inclusive jet $R_{AA}$ and similar observables \cite{Rajagopal:2016uip, Casalderrey-Solana:2018wrw}. Due to the normalization of the quenching weight $D(\varepsilon)$, this uncertainty ends up affecting the whole shape of the distribution.

\subsection{Photon-tagged jet observables}
\label{sec:photonjet_observables}

\subsubsection{Per-photon jet yield}
\label{sec:jet_yield}

The photon-tagged jet yield exhibits a more composite distribution, interpolating between the case when the $\pT$'s of the jet and photon are balanced and when they are not, typically we consider $\pTj > \pTg$. The former is dominated by the leading-order contribution to the hard cross-sections, while the latter cases arise at higher-order processes. The data is presented as a function of the photon-jet $\pT$-balance, defined as $\xjg\equiv\pT/\pTg$, for a given $\pTg$ bin. The jet yield as a function of $\xjg$ can be obtained from the jet yield as a function of $\pT$ from
\begin{equation}
    \left.\frac{\rmd N}{\rmd \xjg}\right|_\text{cuts} \!\!=\!\! \int \rmd\pTg \rmd \pTj \, \delta\left(\xjg - \frac{\pTj}{\pTg} \right) \left.\frac{\rmd N}{\rmd \pTj \rmd \pTg} \right|_{\rm cuts} \,,
\end{equation}
where ``cuts'' refer to additional acceptance cuts on the jet and photon measurements. The experimental data is binned in $\pTg$ and, assuming a constant value of $\pTg$ for each $\pTg$ bin, can be written as
\begin{align}
    \left.\frac{\rmd N}{\rmd \xjg}\right|_\text{cuts}
    &\approx \langle \pTg \rangle_\text{bin} \left.\frac{\rmd N^{pp}}{\rmd \pTj}\right|_{\pTj = \xjg \langle \pTg \rangle_\text{bin}}\,.
\end{align}
In the case of the data used in this study, the $\pTg$ bins are too broad for this approximation to hold. The approximation can be made more precise by adjusting the $\langle\pTg\rangle_\text{bin}$ to each $\xjg$ bin, i.e. $\langle\pTg\rangle_\text{bin} \to \langle\pTg(\xjg)\rangle_\text{bin}$. We have explicitly checked that this method gives a good description of the baseline $pp$ data.

The per-photon jet yield in AA is then obtained by introducing the effect of jet energy loss similarly as in Eq. \eqref{eq:sigma-AA},
\begin{multline}
    \label{eq:sigma-xJ-AA}
    \frac{1}{N_\gamma} \frac{\rmd N^{AA}}{\rmd \xjg} \approx \langle \pTg (\xjg) \rangle_\text{bin} \\
    \times \sum_{i=q,g}\int_0^\infty \rmd \varepsilon \, D_i(\varepsilon)  \left.\frac{1}{N_\gamma}\frac{\rmd \tilde N^{pp}_i}{\rmd \pTj}\right|_{\pTj = \xjg \langle \pTg (\xjg) \rangle_\text{bin} + \varepsilon}\,,
\end{multline}
where $\tilde N^{pp}$ is the baseline jet yield with the replacement of the proton PDFs by nuclear PDFs (nPDFs).

Like for the inclusive jet observables, the $pp$ jet yield is obtained via Monte Carlo event generation, as described in Sec.~\ref{sec:inclusive_observables}. $\tilde N^{pp}$ is obtained by including the nuclear PDF set EPS09 \cite{Eskola:2009uj} at the generator initialization. Jets for which a photon is found back-to-back in the transverse plane, $\Delta\phi>7\pi/8$, are considered. The generated $pp$ per-photon jet yield, $1/N_\gamma\, \rmd N^{pp}/\rmd\xjg$, is shown in Fig.~\ref{fig:pythia_pp}, for the $\pTg$ bins of interest, together with the measured spectrum from ATLAS \cite{ATLAS:2018dgb}. The agreement between the simulation and the data guaranties a correct generation baseline. The nPDF effect is shown in Fig.~\ref{fig:nPDF_effect}. For this observables, we see no significant effects of nuclear modifications to the PDFs.

The factorization ansatz in Eq.~\eqref{eq:fact} is well-established for final-state partons (with positive virtuality).
For consistency, only jets from final hard partons are considered in our analysis of the experimental data. In this sense, the low-momenta regions where the fraction of jets coming from initial state radiation are higher than 20\% are excluded from the analysis.

The generated histograms, for quark- and gluon-initiated jet yields, are fitted by continuous functions. For a fixed $\pTg$ and $\xjg\gtrsim 1$, the per-photon jet yield is found to be steeply fall with $\pTj$, following \eqn{eq:fall_spec}, while for $\xjg\lesssim 1$ the spectrum can be approximated by a linear function. Overall, the jet yield is well described by
\begin{equation}
    \label{eq:balance_spec}
    \frac{1}{N_\gamma}\frac{\rmd \tilde N^{pp}_i}{\rmd \pT} = 
    \begin{cases}
        B^{(i)}_0 + B^{(i)}_1\,\pT\,,\quad \pT\leq\pT^\text{fall} \\
        A^{(i)}_0\left(\frac{\pT^0}{\pT}\right)^{n_i(\pT)}\,,\quad \pT>\pT^\text{fall}
    \end{cases}\,,
\end{equation}
where $\pT^\text{fall}\approx\pTg$ is chosen such that the resulting curve is continuous over $\pT$. Here, $n_i(\pT)$ is truncated at $N=3$. The resulting curves are plotted in App.~\ref{sec:spectra_fits}.

\subsubsection{Photon-tagged jet spectrum and $R_{AA}$}
\label{sec:photon_raa}

We also have considered recent measurements of the nuclear modification factor $R_{AA}$ for photon-tagged jets \cite{ATLAS:2023iad}.
It is defined and described precisely in the same way as the inclusive $R_{AA}$ in Sec.~\ref{sec:inclusive_observables}. The $pp$ spectra are also generated in the same way as for inclusive jets, see Sec.~\ref{sec:inclusive_observables}, with jets for which a photon is found back-to-back in the transverse plane, $\Delta\phi>7\pi/8$, being considered. The resulting curves are shown in the figures of App.~\ref{sec:spectra_fits}. Contrary to the per-photon jet yield, the nuclear modification factor is not binned in $\pTg$. The generated $pp$ spectrum $\sigma^{pp}$, and the respective experimental measurement from ATLAS \cite{ATLAS:2023iad}, are plotted in Fig.~\ref{fig:pythia_pp}, again with a good agreement between generation and data. 

We have not included this data in the main part of the analysis, but will comment on this data thoroughly in Sec.~\ref{sec:other-data} and App.~\ref{sec:photon-tagged}.

\subsubsection{Establishing the baseline for jet energy loss}
\label{sec:photon_raa}

In Fig.~\ref{fig:nPDF_effect} we plot the nPDF effects on the spectrum. In the ratio of $AA$ to $pp$ data, a significant effect is again recovered and it reaches a deviation of 20\% at high $\pT$ for inclusive jets(left panel), and a general deviation also of about 20\% for $\gamma$-jets (lower right panel). It is important to note this base contribution of the nPDF effect for the photon-tagged jet $R_{AA}$ without any final-state energy loss.
The effects of including nPDFs in the analysis are further discussed in App.~\ref{sec:photon-tagged}.

This concludes the effort to establish the baseline spectra. Equations~\eqref{eq:sigma-AA} and \eqref{eq:sigma-xJ-AA} serve as the input for the Bayesian analysis, which we describe in detail now.

\section{Bayesian inference}
\label{sec:bayesian}

\subsection{Inference framework}

Bayesian inference offers a natural way of accessing and dealing with uncertainty. While a frequentist approach gives a point estimate for the parameters $\theta$ of a given model that best describe the observed data $x$, a Bayesian analysis provides us with a full distribution over the model parameters. Within a Bayesian analysis, this distribution over the model parameters represents an update of our prior belief about the parameter distribution due to the observed data. This distribution over the model parameters, given the observed data, is thus called \textit{posterior distribution}, or simply \textit{posterior}, and is given by the Bayes' theorem,
\begin{equation}
\label{eq:Bayes}
    P(\theta|\bm{x}) = \frac{\mathcal{L}(\bm{x}|\theta)P(\theta)}{P(\bm{x})}\,.
\end{equation}
In this formula $P(\theta)$ is the prior distribution, which contains previous beliefs about the model parameters, $\mathcal{L}(\bm{x}|\theta)$ is the \textit{likelihood} of measuring the observed data $x$ given the model parameters $\theta$. Finally, the normalization constant $P(\bm{x}) = \int \rmd \theta \,\mathcal{L}(\bm{x}|\theta)P(\theta)$ is known as the \textit{evidence}, which is interpreted as the probability of the data $\bm{x}$ to be observed. For the likelihood, we use a multivariate normal distribution, which is commonly used by assuming normal distributed uncertainties. The log-likelihood is then given by
\begin{equation}
    \label{eq:loglikelihood}
    \ln[\mathcal{L}(\bm{x}|\theta)] = -\frac{1}{2}\ln\left[(2\pi)^n\det\Sigma\right] -\frac{1}{2}\Delta \bm{x}^T\Sigma^{-1}\Delta \bm{x} \,,
\end{equation}
where $n$ is the number of data points used for inference, $\Sigma$ is a covariance matrix that correlates experimental and model uncertainties, and $\Delta\bm{x} = \bm{x}_\theta-\bm{x}$, with $\bm{x}_\theta$ being the modeled value for $\bm{x}$.

The posterior distribution can be used to infer new unobserved data points by computing the respective \textit{posterior predictive distribution}, i.e. the probability distribution of the new data points. The posterior predictive distribution is computed by marginalizing the likelihood of the new data over the posterior distribution
\begin{equation}
\label{eq:post_predictive}
    P(\bm{x}_\text{new}|\bm{x})=\int \rmd \theta \,\mathcal{L}\left(\bm{x}_\text{new}|\theta\right)\,P\left(\theta|\bm{x}\right)\,.
\end{equation}
The posterior distribution of another variable of interest $\psi$, which is not a model parameter, can also be obtained by marginalized the joint probability distribution $P(\psi,\theta|\bm{x})$ of the variable $\psi$ and the model parameters $\theta$ over the posterior distribution. In the case $\psi$ is a function of the model parameters $\theta$, i.e. $\psi=\psi(\theta)$, then $P(\psi,\theta|\bm{x})=\delta\big(\psi-\psi(\theta)\big) P(\theta|\bm{x})$, and
\begin{equation}
\label{eq:post_newvar}
    P(\psi|\bm{x})=\int \rmd \theta\,\delta(\psi-\psi(\theta))\,P(\theta|\bm{x}) \,.
\end{equation}
The prior distribution $P(\psi)$ of the variable $\psi$ can also be computed, by simply switching the posterior $P(\theta|\bm{x})$ in Eq.~\ref{eq:post_newvar} with the prior $P(\theta)$.

The posterior distribution contains all the information about the inferred model parameters, and respective uncertainty. From it, we can extract quantities such as credible intervals. A particular type of credible interval is the highest density interval (HDI), defined as the narrowest interval of the probability distribution that contains a specified probability mass. Mathematically, the $\alpha$\% HDI of a probability distribution $p(\theta)$ is defined as
\begin{equation}
    \label{eq:hdi}
    \underset{\theta_\text{min}, \theta_\text{max}}{\arg\min}\,\theta_\text{max}-\theta_\text{min}\,:\,\frac{\alpha}{100}=\int_{\theta_\text{min}} ^{\theta_\text{max}} \rmd \theta\, p(\theta)\,.
\end{equation}
Single value estimates can also be computed, such as the maximum a posteriori (MAP) estimate, and the moments of the posterior distribution. The MAP estimate, which corresponds to the mode of the distribution, i.e. its most probable value, is defined as
\begin{equation}
\label{eq:MAP}
    \theta_\text{\tiny MAP} = \underset{\theta}{\arg\max}\,P(\theta|\bm{x})\,.
\end{equation}
The $n^{\rm th}$ moment of the posterior distribution is defined as
\begin{equation}
\label{eq:moments}
    \langle\theta^n\rangle = \int\rmd\theta\,\theta^n\,P(\theta|\bm{x})\,.
\end{equation}
In particular, the expectation value $\langle\theta\rangle$, which corresponds to the mean of the distribution, is obtained for $n=1$.

The posterior distribution can be computed numerically via a Markov Chain Monte Carlo (MCMC) algorithm. MCMC refers to a class of algorithms that access the posterior by directly sampling it, without having to rely on the challenging, and sometimes intractable, computation of the evidence $P(\bm{x})$ in Eq.~\eqref{eq:Bayes}. The resulting posterior samples offer an easy way to compute credible intervals, MAP estimates, and moments of the distribution, e.g. expectation values, variances and co-variances. An overview of these class of algorithms and examples of their use for Bayesian analysis can be found in \cite{Andrieu2003,Albert2019}. An introduction to Bayesian inference in physics can be found in \cite{RevModPhys.83.943} and, more specifically, for heavy-ion collisions in \cite{Paquet:2023rfd}. For a more detailed introduction to Bayesian inference, see e.g. \cite{Albert2019,DAgostini:2003bpu}.

In this work, our implementation of Bayesian inference closely follows the implementation from the JETSCAPE framework used in \cite{JETSCAPE:2020mzn,JETSCAPE:2021ehl, JETSCAPE:2023ikg, JETSCAPE:2024cqe}. Parallel tempering is used as the MCMC algorithm, where 100 Markov chains are generated with a burn-in period of 500 samples of the posterior. The burn-in samples are discarded to remove any dependence on the starting configuration. We have explicitly checked that the algorithm reaches an equilibrium state within the assigned burn-in period, and that a longer burn-in period does not improve our results. After the burn-in period, 5000 samples are collected, giving a total of $5\times10^5$ draws from the posterior distribution available for analysis.

\subsection{Physical model emulator}

Equations~\eqref{eq:sigma-AA} and \eqref{eq:sigma-xJ-AA} constitute the model for the inclusive jet and $\gamma$-jet observables, respectively. The model implements quark-gluon fractions from the outset. Therefore, we have in total 4 model parameters, two for each jet-initiating parton flavor for each of the parametrizations of the quenching weight $D(\varepsilon)$ described in Sec.~\ref{sec:parametrizations}.

In MCMC, at each step of the posterior sampling, the model has to be evaluated at the sampled values in the parameter space. This poses a numerical problem, since the precise numerical computation of the integral in Eqs.~\eqref{eq:sigma-AA} and~\eqref{eq:sigma-xJ-AA} would represent either an impractical expense in computation time or a significant decrease in the precision of the integration. Moreover, these would be further enhanced as the dimensionality of the problem increases, with the possible addition of more model parameters and the increase of the data set used for the inference foreseen for future applications.

This problem is addressed by introducing a physical model emulator that works as a statistical surrogate of the model in study. The implemented model emulator uses Gaussian Processes (GP), a stochastic process that is used to model unknown functions. With GP, the model can be interpolated over the parameter space. The use of an emulator trained on previously calculated model outputs, makes the model calculation at each sampling step of the MCMC algorithm becomes much more efficient. A more complete definition and explanation of GP can be found in \cite{williams2006gaussian,bishop2006pattern,Rasmussen2005}. The use of GP for model emulation in heavy-ion collisions is studied in \cite{Weiss:2023yoj}. The model emulator used in this work was first introduced for Bayesian inference in the context of heavy-ion collisions in \cite{Novak:2013bqa}. Our implementation follows the one in the JETSCAPE framework, see \cite{JETSCAPE:2020mzn,JETSCAPE:2021ehl, JETSCAPE:2023ikg, JETSCAPE:2024cqe}, and is outlined below. Further work that relies on the use of a similar emulator to constrain QGP properties can be found in e.g. \cite{Paquet:2023rfd, Bernhard:2015hxa, Bernhard:2019bmu, Auvinen:2017fjw, Nijs:2020roc}.

The model output, i.e. the data points, is numerically calculated at a number of randomly chosen points in the parameter space, called the design points. The calculated model outputs are then used as points to be interpolated in a multidimensional hyper-space. To decrease the dimensionality of the space to be interpolated, i.e. the total number of data points, dimensionality reduction is performed via Principal Component Analysis (PCA). The use of PCA is justified by the high level of correlated information across the data points. Therefore, the major part of the model information is carried by a small subset of the full model output. Within this work, we keep the first 8 dominant principal components (PCs), which is enough to explain 99\% of the variance, indeed showing the high correlation across the model outputs. Each of the resulting PCs is then independently interpolated by a GP. After training the GP as an interpolation of each PC, its value can be computed for any point in the parameter space, and the complete model output can be reconstructed via inverse PCA transformation. Once the full physical model emulator is trained on the design points, the emulator can be used as a non-parametric map between a point in the model parameter space and the data points given as the model output. There, the uncertainties stemming of each step in this process are also addressed. We have explicitly checked that the uncertainties introduced to speed-up the inference, through dimensionality reduction and emulation, do not affect the quality of our results.

\subsection{Closure tests}
\label{sec:closure_tests}

\begin{figure}
    \centering
    \includegraphics[width=0.9\columnwidth]{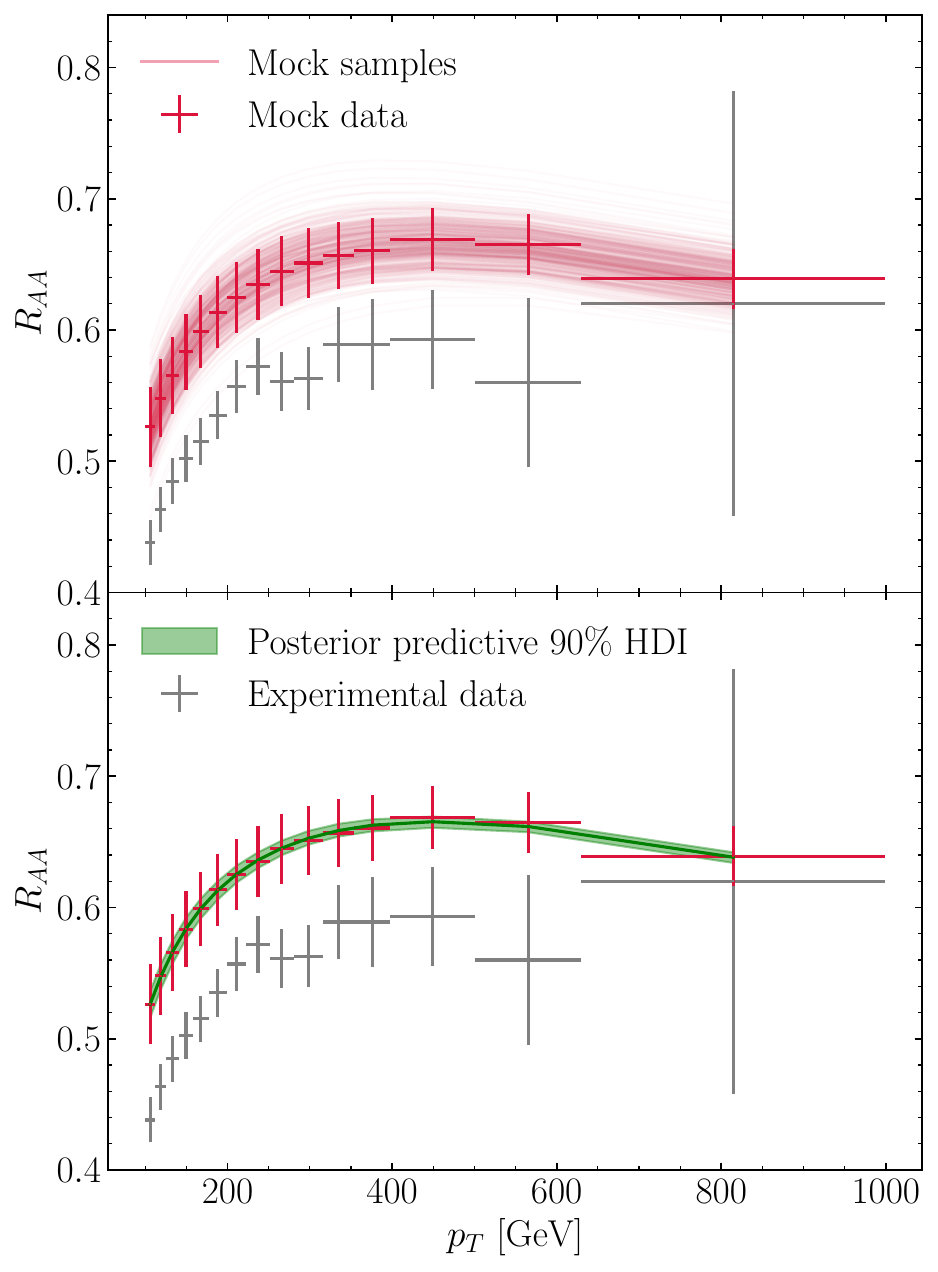}
    \caption{An example on how the closure tests are performed used the described Bayesian inference setup for the inclusive $R_{AA}$. In the top panel, generated replica samples (in red lines) and the mock data (in red points). The 90\% HDI of the posterior predictive distribution of the $R_{AA}$, obtained with the inference on the mock data, is shown in the bottom panel (green lines). Experimental data, from \cite{ATLAS:2018gwx}, is plotted in both panels for visual reference (gray points).}
    \label{fig:ct_example}
\end{figure}

\begin{figure*}
    \centering
    \includegraphics[width=0.9\textwidth]{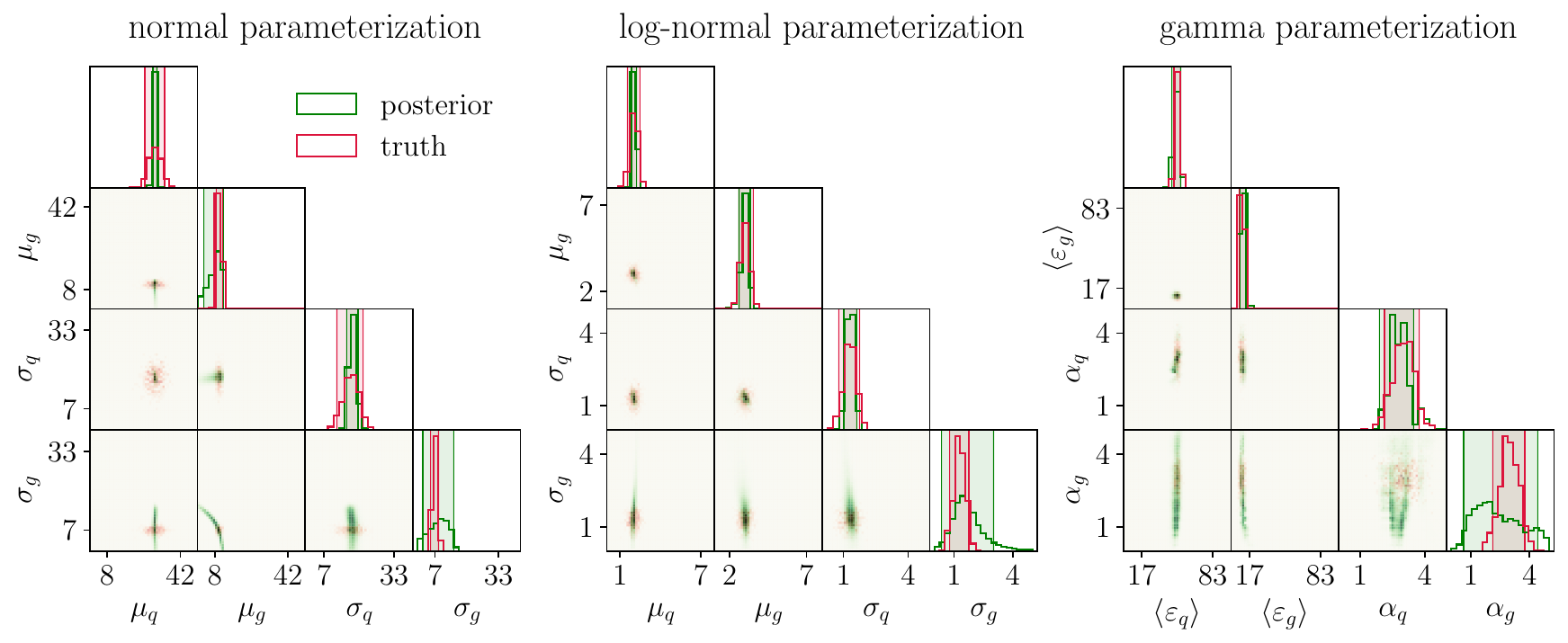}
    \caption{Posterior distributions obtained for the closure tests presented in the form of a corner plot. The posteriors for the three studied energy-loss distribution parametrizations are compared with the true distributions used to generate the mock data. The 90\% HDI is shown, as vertical lines, to help the comparison.}
    \label{fig:ct_corner}
\end{figure*}

Closure tests were performed to ensure the validity and consistency of the inference setup described above. Such tests consist on performing a Bayesian analysis using artificial data, so-called replicas (alternatively as ``mock'' data or pseudo-data). Replicas are generated by computing the model data points for a chosen value of the model parameters. Next, a full Bayesian analysis is done, and the resulting posterior distribution is compared with the true parameter distribution, i.e. the distribution used to generate the data in the first place. Ideally, the posterior should be peaked closely to the true value of the chosen parameters, but this depends both on the uncertainties of the emulator and on the sensitivity of the observables to the parameters. Metrics quantifying the success of a closure tests have been discussed in \cite{Weiss:2023yoj,JETSCAPE:2021ehl,JETSCAPE:2023ikg}.

In practice, the replicas are generated by numerically computing the full model for a set of random points following a normal distribution in the parameter space, with a chosen mean and standard deviation. This procedure creates a set of values for each data point. The mean of the produced set gives a replica data point, with an uncertainty given by the respective standard deviation.

For the Bayesian analysis, a flat prior is used for the model parameters, meaning that the design points of the physical model emulator are drawn from a uniform distribution within the desired range. The range is chosen to be wide enough to explore the preferred values. Bayesian inference is then performed on the generated replica data, using the framework described above to obtain the posterior distributions of the model parameters. To visually verify the quality of the inference, the 90\% HDI of the posterior predictive distribution can be calculated using Eq. \eqref{eq:hdi}, and compared with the replica data points.

To serve as an example, Fig.~\ref{fig:ct_example} shows the method just described for the parametrization of energy loss using a normal distribution. In the left plot of Fig.~\ref{fig:ct_example} we show the generated data (red lines) and the resulting set of data points (red crosses) computed for a specific observable, the inclusive $R_{AA}$. The experimental data is also presented \cite{ATLAS:2018gwx}, as a visual reference (gray crosses). The replicated data have been chosen with a specific offset with respect to the experimental data. On the right plot of Fig.~\ref{fig:ct_example}, the 90\% HDI of the obtained posterior predictive distributions for the replicated data points.

Closure tests were performed for the three energy loss parametrizations under study, described in Sec.~\ref{sec:parametrizations}. In Fig.~\ref{fig:ct_corner} we show the obtained posterior distributions, in the form of corner plots, together with the true parameter distribution (referred to as `truth' with open histograms in this figure), i.e the parameter distribution used for generating the replicated data. The 90\% HDI is also shown to help the comparison. A good agreement between the posterior and the true distribution is in general achieved. This demonstration serves to strengthen the confidence for extracting meaningful information from data on this particular observable.

\begin{figure}
    \centering
    \includegraphics[width=0.9\columnwidth]{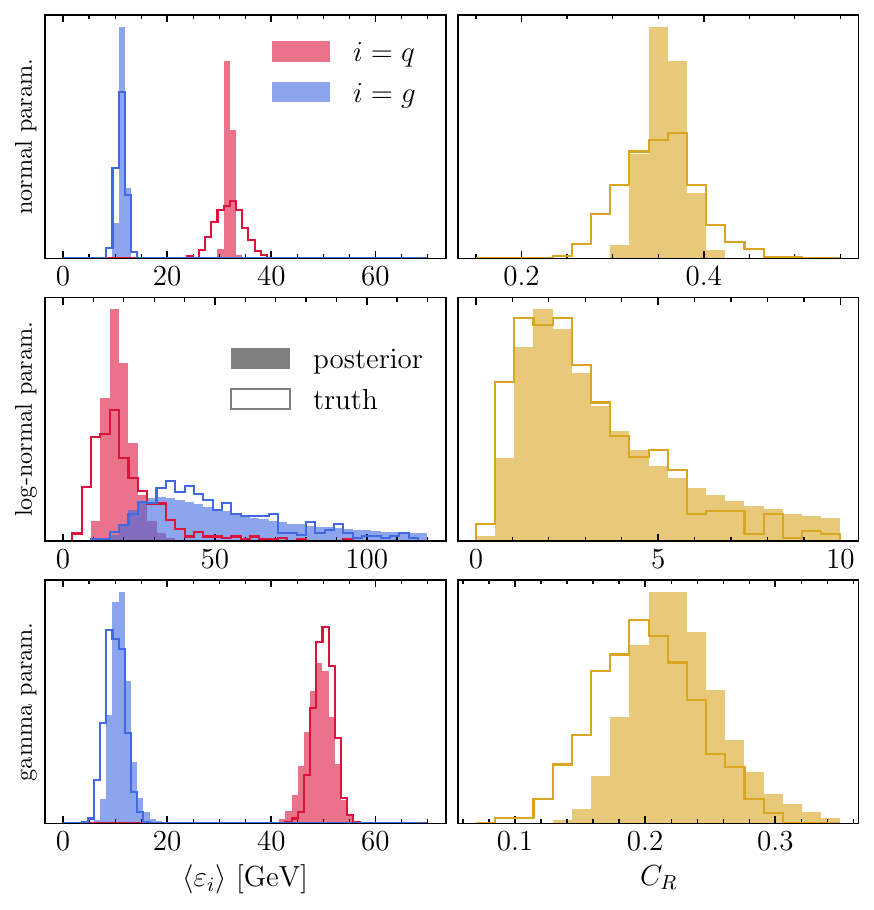}
    \caption{Prior and posterior distributions of the mean energy loss by the quark- and gluon-initiated jets obtained for the closure tests, for the three studied distributions, obtained from marginalizing the posteriors. The color ratio is also shown.}
    \label{fig:mel_ct}
\end{figure}

Going deeper into the results of the closure tests, we also check whether the shapes and moments are commensurate with the generated samples. The posterior distributions for the mean quark/gluon energy loss $\langle\varepsilon_i\rangle$, i.e. $P(\langle\varepsilon_i\rangle|\bm{x})$, can separately be calculated for quark- ($i=q$) and gluon-initiated ($i=g$) jets from Eq. \eqref{eq:post_newvar} using the corresponding posterior distributions for the three energy-loss parametrizations.
In the same way the color ratio $\langle \varepsilon_g \rangle/\langle \varepsilon_q \rangle$, i.e. $P(C_R|\bm{x})$, can be obtained. Figure~\ref{fig:mel_ct} shows both the posterior and true distributions obtained for the quark/gluon energy loss and respective color ratios, for the three energy-loss parametrizations in study. A match between the posterior distributions and the true ones is in general achieved for the three parametrizations, validating the used framework. To comment further on the right column of Fig.~\ref{fig:mel_ct}, we have not assumed a universal value for the color ratio across the three parametrizations. We currently turn to the analysis of real experimental data.

\section{Inference on LHC measurements}
\label{sec:inference}

We finally turn to the main task of our work, which is to extract information about jet energy-loss directly from experimental data in a data-driven fashion. Bayesian inference is performed on the experimental data, see also Sec.~\ref{sec:data}: the inclusive $R_{AA}$, and respective relative $R_{AA}$ for different rapidity bins; and the per-photon yield of photon-tagged jets for different $\pTg$ bins. The inference is performed using the same procedure described in Sec.~\ref{sec:closure_tests} for the closure tests, for the three energy-loss distribution parametrizations described in Sec.~\ref{sec:parametrizations}.

\subsection{Study of universality}
\label{sec:universality}

The universality of the energy-loss processes assumes that the same quenching weight can be applied (under assumptions expounded in detail above and on average) no matter the underlying hard process. We will test this assumption by comparing the inclusive jet and the photon-tagged jet data sets. To study the universality of jet energy-loss across different observables, two analyses are done in a first stage of the study: 
\begin{description}
    \item[Analysis A] Inference is performed using only the inclusive jet observables, and the posterior distributions are used to predict the photon-tagged jet observables. The following observables are used for inference: inclusive $R_{AA}$; inclusive $R_{AA}^y/R_{AA}^{y<0.3}$ for 5 rapidity bins.
    \item[Analysis B] Inference is performed using only photon-tagged jet observables, and the posterior distributions are used to predict inclusive jet observables. The following observables are used for inference: photon-tagged multiplicity for 4 $p_T^\gamma$ bins.
\end{description}
Finding internal consistency among these two analyses is taken as evidence for universal features of the energy-loss process (under assumptions).

\begin{figure*}
    \centering
    \includegraphics[width=0.9\textwidth]{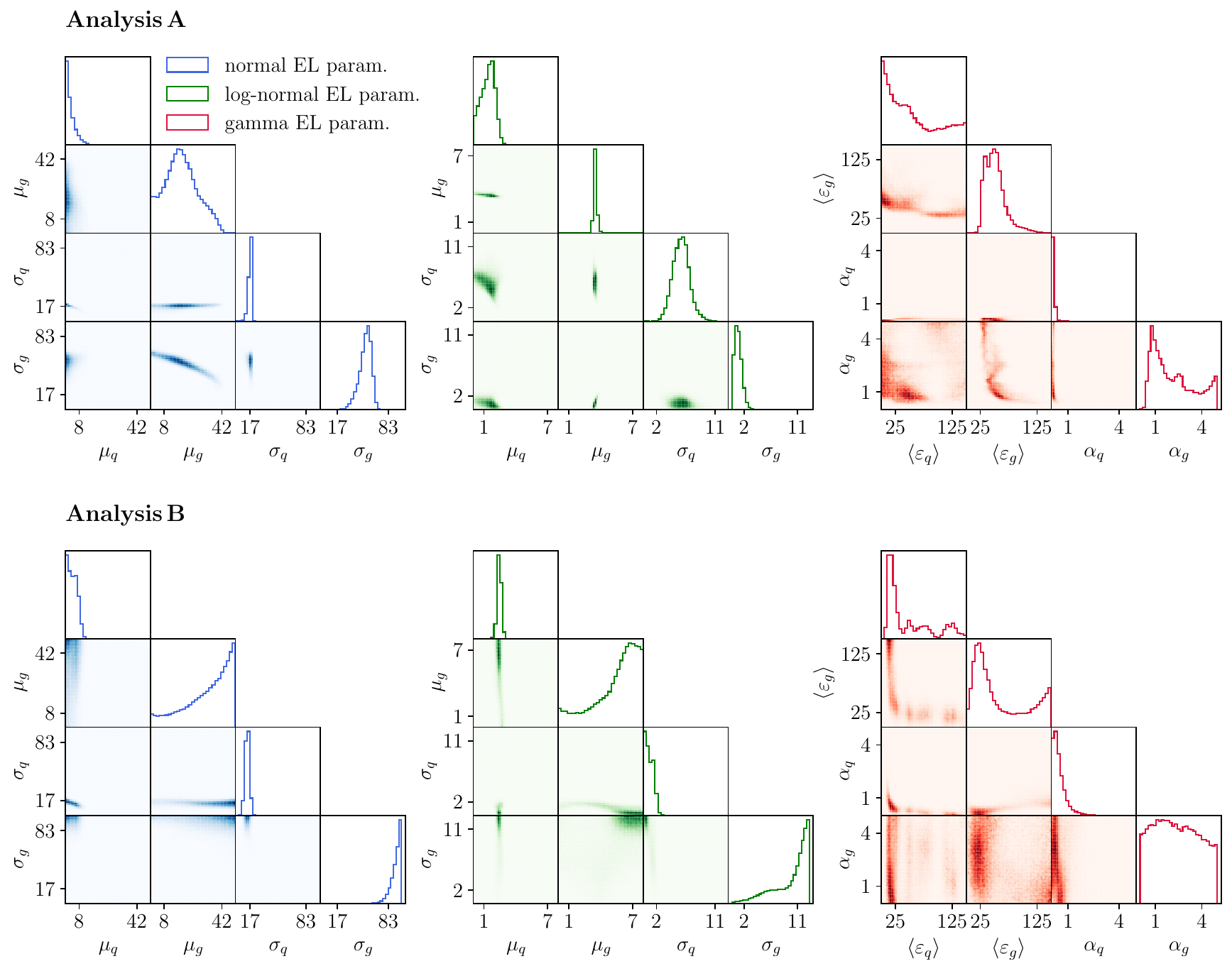}
    \caption{Posterior distributions and respective correlations, in the form of a corner plot, for the three energy-loss parametrizations: normal, in blue; log-normal, in green; and gamma, in red. The posteriors are shown for the two Bayesian analyses: analysis A, using only inclusive jet observables for inference; and analysis B, using only photon-tagged jet observables.}
    \label{fig:corner_ab}
\end{figure*}
Figure~\ref{fig:corner_ab} summarizes the posterior distributions obtained for analyses A and B as corner plots. For the three energy-loss parametrizations, the posterior is more constrained in analysis A, when inference is done using inclusive jet observables, compared to analysis B, which only uses photon-jet observables. The higher constraining power from the inclusive jet observables shows that these observables contain more information about the energy loss of quark- and gluon-initiated jets compared with the photon-tagged jet observables, which are dominated by quark-initiated jets, see Fig.~\ref{fig:qrk_frac}. This is in line with what is expected from the discussion in Sec.~\ref{sec:data} regarding the sensitivity to quark-jet versus gluon-jet quenching for the different observables. The effect of the jet flavor of the photon-tagged jets is more pronounced in analysis B, where the photon-tagged jet observables show a lack of constraining power on the posterior distribution of the model parameters related with the gluon-initiated jet energy loss. Regarding the different parametrizations, we see that the normal and log-normal are better constrained by the data.


As a first, important step of our analysis, we would like to critically examine the main assumption of our setup, namely the factorization between the hard-processes and the subsequent quenching phenomena, as stated in Eq.~\eqref{eq:fact} and further discussed in Secs.~\ref{sec:intro} and \ref{sec:theory}. This assumption states that, for similar conditions of the underlying medium (which, in our context, reduces to considering the same centrality class and center of mass energy) and for the same class of reconstructed jets (meaning here, the same cone-angle $R$), the same energy-loss distribution can be used for different jet observables. As a first goal, we wish to establish the degree to which, under minimal assumptions, the existing data support such a picture.

\begin{figure*}
    \centering
    \includegraphics[width=\textwidth]{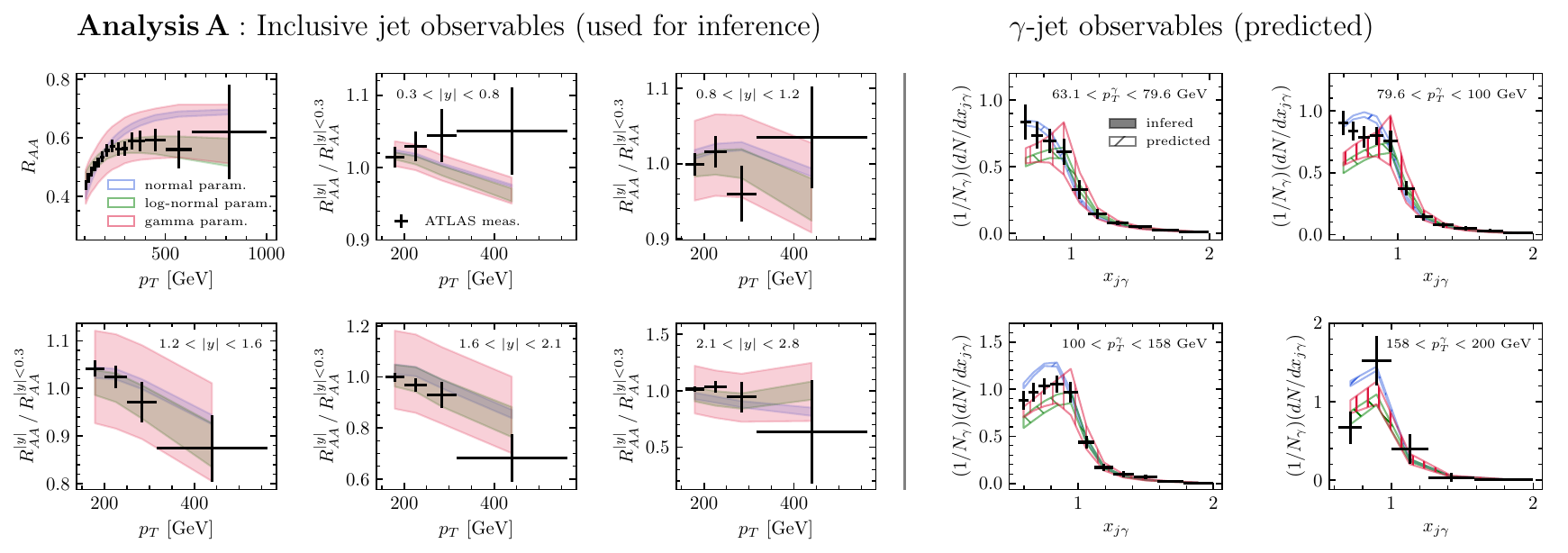}\\
    \includegraphics[width=\textwidth]{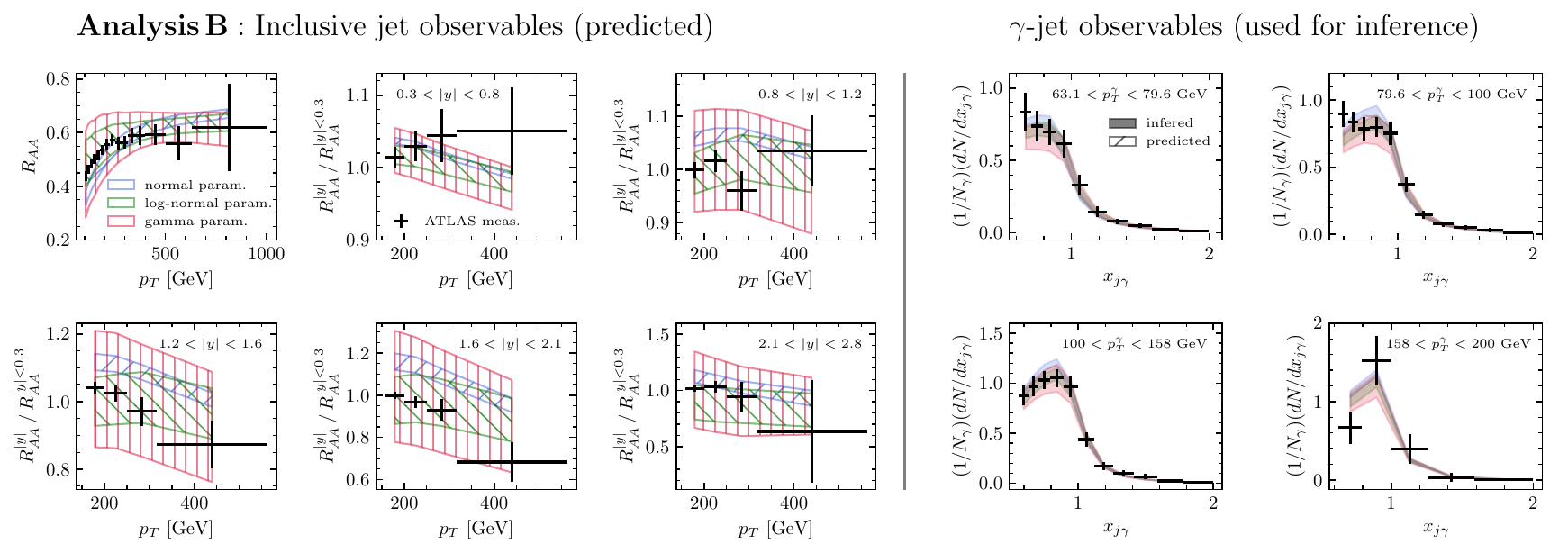}
    \caption{90\% HDI of the posterior predicative distributions obtained for analysis A (upper panels) and analysis B (lower panels) for all the observables in the study. Observables used for inference are marked with filled bands  while the textured bands indicate predicted observables.}
    \label{fig:fit_pred_a}
\end{figure*}
Once the Bayesian inference is done for both analysis A and B, the posterior predictive distributions are computed for observables that were not used for inference, in this way constituting predictions for unobserved data. Posterior predictive distributions are also computed for the the observables that were used for inference, in order to check the quality of the inference on the observed data. Figure~\ref{fig:fit_pred_a} show the 90\% HDI of the posterior predictive distributions for all the observables in study, for both analysis A (upper panels) and B (lower panels), for the three energy-loss parametrizations (normal in blue bands, log-normal in green bands and gamma in red bands). Focusing on the ability to reproduce data used in the inference (filled bands in Fig.~\ref{fig:fit_pred_a}), a good agreement with the observed data is in general obtained which is consistent across the different parametrizations. In more detail, the normal distribution is in tension with the high-$\pT$ data on the inclusive $R_{AA}$ at mid-rapidity, see Fig.~\ref{fig:fit_pred_a} (upper panels, upper-leftmost plot). No such tension is, however, observed for the log-normal and gamma distributions.

When looking into the prediction of the unobserved data (represented by textured bands in Fig.~\ref{fig:fit_pred_a}), the difference between the analysis A and B becomes evident: without the constraining power of the photon-tagged jet observables over the the gluon energy-loss parameters, the capability of predicting other observables that contain a significant fraction of gluon-initiated jets, such as the inclusive jet observables, is substantially decreased. On the contrary, a very good agreement between the predicted photon-tagged jet observables and the data is achieved in analysis A. This indicates that the inclusive data on jet suppression and, crucially, including its rapidity dependence, has enough constraining power to disentangle the quark and gluon contributions to the overall energy loss. In contrast, the photon-jet data on their own are dominated by quark-jets, and fails to adequately constrain the gluon component of jet energy loss.

These results demonstrate clear evidence of universality: when performing inference on data with constraining power over the energy loss distributions of both jet-initiating partons, the model is capable of describing other unobserved observables with different ratios of quark-/gluon-initiated jets. It supports the notion that the energy-loss distribution is largely an effect of the medium and has only a weak dependence on jet properties\,\textemdash\, at least, as encoded in the analyzed data.

\subsection{Global analysis: retrieving physical insight}
\label{sec:insight}

\begin{figure*}
    \centering
    \includegraphics[width=0.9\textwidth]{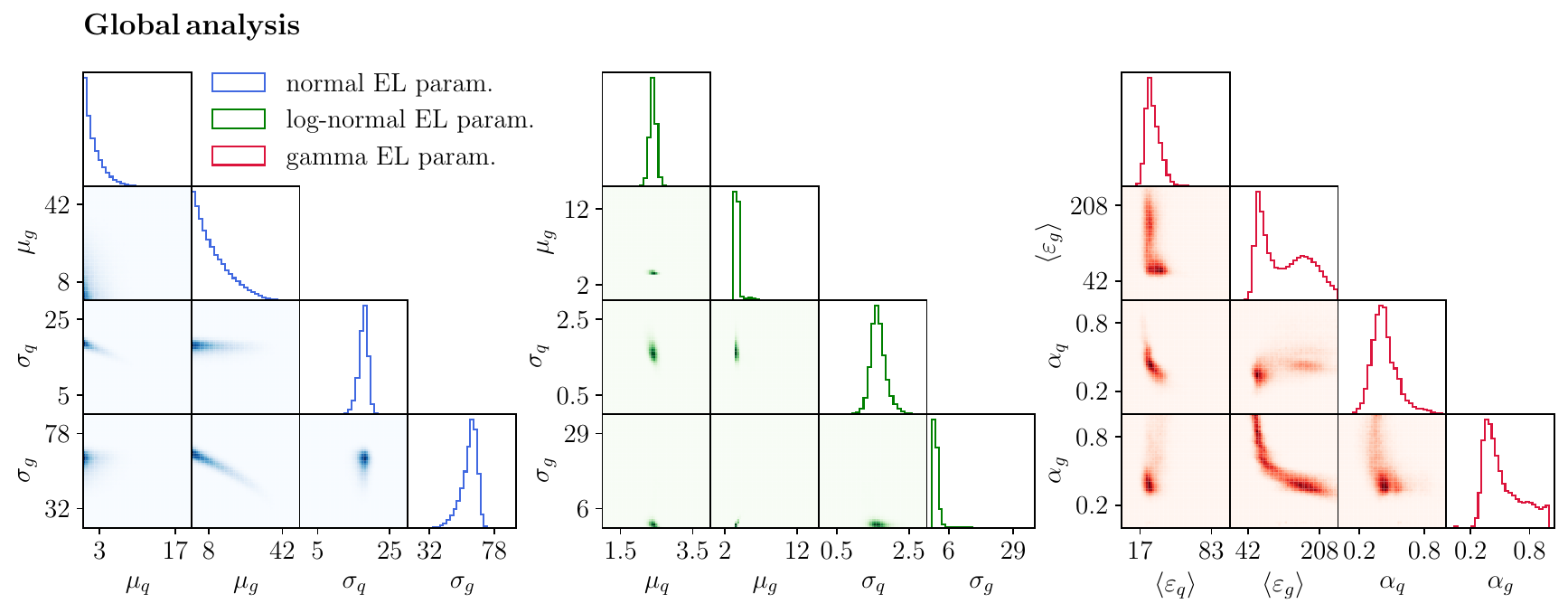}
    \caption{Posterior distributions and respective correlations, in the form of a corner plot, for the three energy-loss parametrizations: normal, in blue; log-normal, in green; and gamma, in red. The posteriors are shown for the global Bayesian analyses, where all the measurements in study were used for inference.}
    \label{fig:corner_global}
\end{figure*}

In order to maximize the amount of information retrieved directly from the experimental data, a Bayesian analysis is carried out using simultaneously all the observables discussed above. The inference procedure is the same as in the analyses in sections \ref{sec:closure_tests} and \ref{sec:universality}. The resulting posterior distributions are shown as corner plots in Fig.~\ref{fig:corner_global}, for the three energy-loss parametrizations. From a broad perspective, the inference across all available data has constrained the extracted parameters to a narrower range and reduced their correlations. We put off a further discussion about the obtained parameter values for Sec.~\ref{sec:conclusions}.

\begin{figure*}
    \centering
    \includegraphics[width=\textwidth]{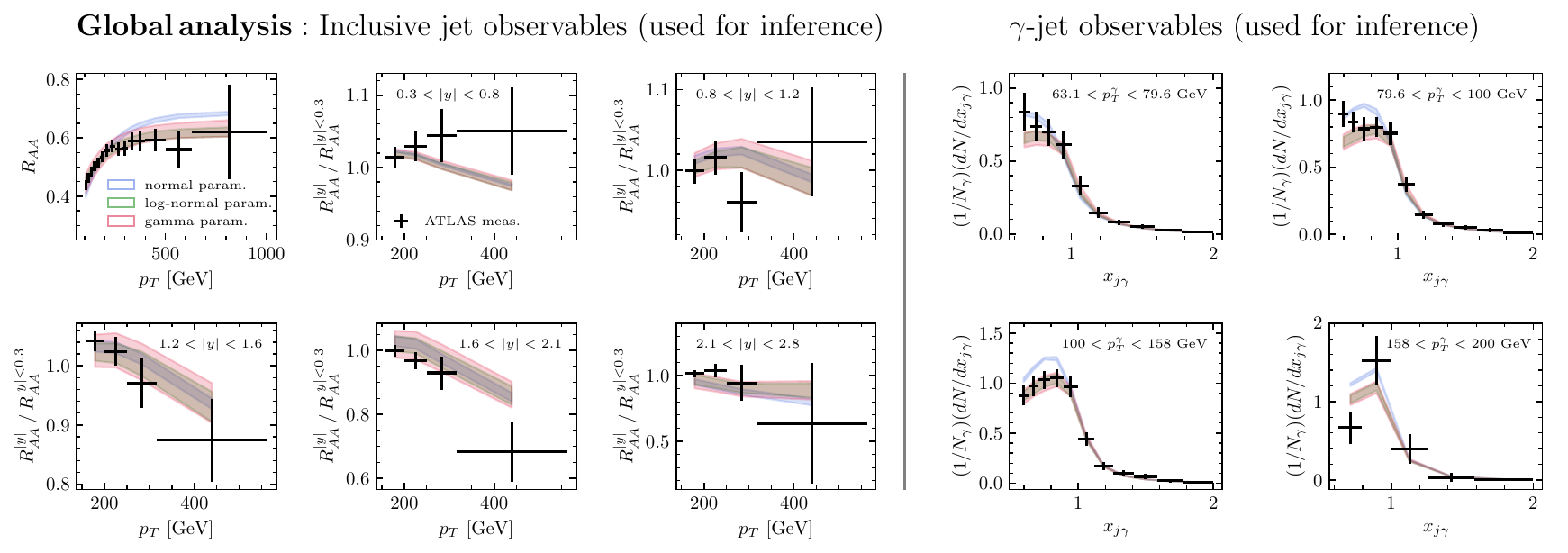}
    \caption{90\% HDI of the posterior predicative distributions obtained for the global analysis, for all the observables in the study.}
    \label{fig:fit_pred_c}
\end{figure*}

The obtained posterior predictive distributions for all the observables in study is shown in Fig.~\ref{fig:fit_pred_c}. Compared to the inference on parts of the data in Sec.~\ref{sec:universality} the 90\% bands are now much narrower. This can be attributed mostly to the very precise data-points for the inclusive $R_{AA}$ at mid-rapidity in the $100 \text{ GeV}< \pT < 300 \text{ GeV}$ range. A good agreement with the data is achieved, showing that the data can be described by any of the three parametrizations, see also Tab.~\ref{tab:chi2} (top row). The normal parametrization is the one that perhaps performs narrowly the worst, again with the high-$\pT$ behavior of the $R_{AA}$ at mid-rapidity that sticks out. These features indicate that the quark and gluon contribution to jet quenching is well constrained by the experimental data.

\subsubsection{Leave-one-out study on the constraining power of each observable}
\label{sec:leave-one-out}

The leave-one-out method, widely used in machine learning approaches to validate a specific model, can be used here to assess the constraining power of each individual observable. In this example, inference is performed on all observables but one, and the posterior predictive distribution is computed for both the observed data and the unobserved data, which correspond to the observable left out. We evaluate the constraining power of the observable left out by computing the expectation value and 90\% HDI for the reduced chi-square $\tilde\chi^2$ of the posterior predictive distributions, including both the observed and unobserved data. For a given set of model parameters $\theta$, the reduced chi-square is defined as
\begin{equation}
    \label{eq:chi-square}
    \tilde\chi^2(\theta)=\frac{(\bm{x}-\bm{x}_\theta)^2}{\sigma_{\bm{x}}^2} \big/ n_{dof.} \,,
\end{equation}
where $\bm{x}$ and $\sigma_{\bm{x}}$ are the measured data points and respective uncertainties, $\bm{x}_\theta$ are the modeled value for $\bm{x}$ for the given $\theta$, and $n_{dof.}$ is the number of degrees of freedom. In this case $n_{dof.}=n-4$, with $n$ being the number of data points and 4 being the number of inferred model parameters. A posterior distribution, and further the expectation value and 90\% HDI, can be computed used Eqs.~\eqref{eq:post_newvar} and \eqref{eq:hdi}. This procedure is repeated for each observable, and the resulting reduced chi-squares are compared.

Table~\ref{tab:chi2} shows the reduced chi-squares (expectation value and 90\% HDI) obtained when leaving out one observable at a time for the three energy-loss parametrizations. The reduced chi-square obtained when no observable is left out is shown as a reference. For all the three parametrizations, leaving out the inclusive jet $R_{AA}$ substantially increases the reduced chi-square. When leaving the $R_{AA}$ out, we also notice a significant widening of the 90\% HDI. We conclude that excluding the $R_{AA}$ reduces the constraining power of the data used for inference. This highlights $R_{AA}$ as the most constraining observable, followed by the relative $R_{AA}$ at the two most forward rapidities.
Leaving out any other single observable constitutes no major change to the constraining power of the remaining observables.

\begin{table}
	\centering
    \begin{tabular}{lcccc}
        & & \multicolumn{3}{c}{EL parametrization} \\
		\cmidrule{3-5}
        observable left out & $\quad$ & normal  & log-normal & gamma \\ \midrule\midrule
        none & & $2.28^{+0.06}_{-0.07}$ & $1.41^{+0.12}_{-0.11}$ & $1.70^{+0.36}_{-0.29}$ \\ \midrule
        $R_{AA}$ & & $2.77^{+0.70}_{-0.54}$ & $2.09^{+1.01}_{-0.78}$ & $2.06^{+0.94}_{-0.65}$ \\[6pt]
        $R_{AA}^{|y|}\,/\,R_{AA}^{|y|<0.3}$ & & $2.29^{+0.07}_{-0.07}$ & $1.44^{+0.14}_{-0.13}$ & $1.69^{+0.35}_{-0.29}$ \\[-1pt]
        {\scriptsize $0.3<|y|<0.8$} & & & & \\[6pt]
        $R_{AA}^{|y|}\,/\,R_{AA}^{|y|<0.3}$ & & $2.29^{+0.07}_{-0.07}$ & $1.46^{+0.15}_{-0.15}$ & $1.71^{+0.38}_{-0.31}$ \\[-1pt]
        {\scriptsize $0.8<|y|<1.2$} & & & & \\[6pt]
        $R_{AA}^{|y|}\,/\,R_{AA}^{|y|<0.3}$ & & $2.29^{+0.07}_{-0.08}$ & $1.48^{+0.16}_{-0.17}$ & $1.75^{+0.38}_{-0.35}$ \\[-1pt]
        {\scriptsize $1.2<|y|<1.6$} & & & & \\[6pt]
        $R_{AA}^{|y|}\,/\,R_{AA}^{|y|<0.3}$ & & $2.44^{+0.18}_{-0.19}$ & $1.59^{+0.25}_{-0.28}$ & $1.85^{+0.55}_{-0.44}$ \\[-1pt]
        {\scriptsize $1.6<|y|<2.1$} & & & & \\[6pt]
        $R_{AA}^{|y|}\,/\,R_{AA}^{|y|<0.3}$ & & $2.38^{+0.15}_{-0.15}$ & $1.58^{+0.23}_{-0.25}$ & $1.79^{+0.45}_{-0.37}$ \\[-1pt]
        {\scriptsize $2.1<|y|<2.8$} & & & & \\[6pt]
        $N_\gamma^{-1}\cdot \rmd N/\rmd \xjg$ & & $2.29^{+0.07}_{-0.07}$ & $1.42^{+0.11}_{-0.11}$ & $1.67^{+0.34}_{-0.27}$ \\[-1pt]
        {\scriptsize $63.1<p_T^\gamma<79.6$ GeV} & & & & \\[6pt]
        $N_\gamma^{-1}\cdot \rmd N/\rmd \xjg$ & & $2.30^{+0.08}_{-0.08}$ & $1.44^{+0.13}_{-0.13}$ & $1.73^{+0.33}_{-0.32}$ \\[-1pt]
        {\scriptsize $79.6<p_T^\gamma<100$ GeV} & & & & \\[6pt]
        $N_\gamma^{-1}\cdot \rmd N/\rmd \xjg$ & & $2.30^{+0.08}_{-0.08}$ & $1.43^{+0.13}_{-0.12}$ & $1.72^{+0.40}_{-0.32}$ \\[-1pt]
        {\scriptsize $100<p_T^\gamma<158$ GeV} & & & & \\[6pt]
        $N_\gamma^{-1}\cdot \rmd N/\rmd \xjg$ & & $2.29^{+0.07}_{-0.07}$ & $1.41^{+0.11}_{-0.11}$ & $1.67^{+0.34}_{-0.27}$ \\[-1pt]
        {\scriptsize $158<p_T^\gamma<200$ GeV} & & & & \\ \bottomrule                   
	\end{tabular}
	\caption{Reduced chi-square resulting form leave-one-out validation analysis. The chi-square presented correspond to the obtained reduced chi-square when the observable on the left column is not used for inference. Values are compared with the chi-square values obtained when all the observables are used for inference (first row).}
	\label{tab:chi2}
\end{table}

\subsubsection{Mean jet energy-loss and color dependence}

\begin{figure}
    \centering
    \includegraphics[width=0.9\columnwidth]{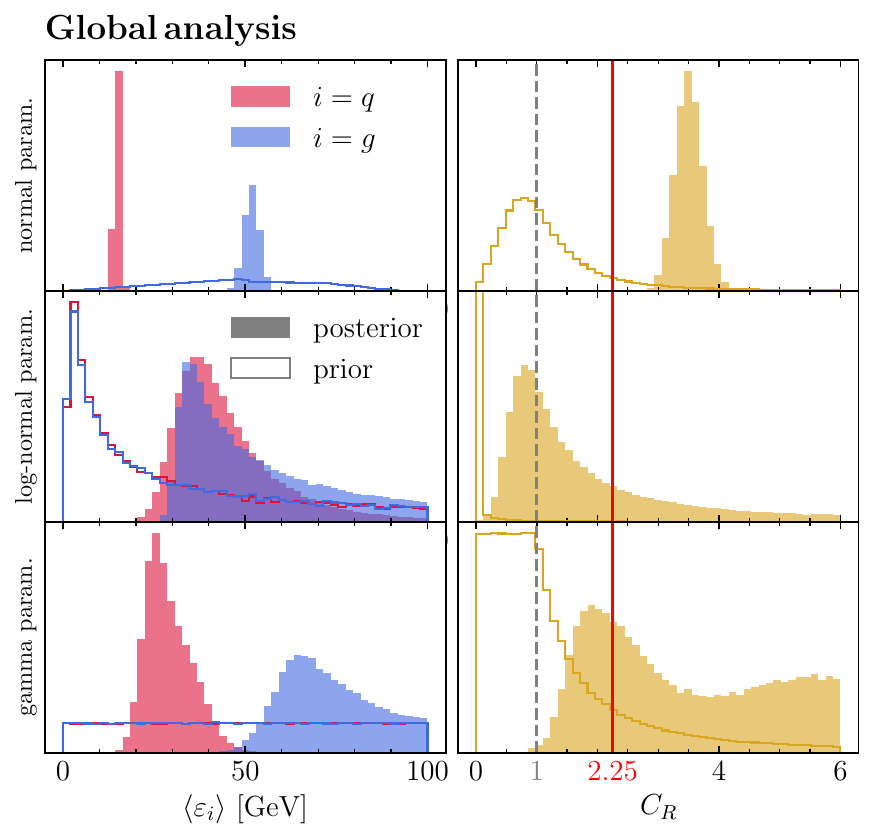}
    \caption{Prior and posterior distributions of the mean energy loss by the quark- and gluon-initiated jets, for the three studied distributions, obtained from marginalizing the posteriors (left column). The color ratio is also shown (right column).}
    \label{fig:mel}
\end{figure}

The posterior distribution of the mean jet energy-loss $P(\langle\varepsilon_i\rangle|\bm{x})$ quark- and gluon-initiated jets and the posterior distribution of the color ratio $P(C_R|\bm{x})$ can be computed for all three parametrizations as described in Sec.~\ref{sec:closure_tests}. Figure~\ref{fig:mel} shows the posterior distributions for the mean jet energy-loss (left column), and the respective color ratio $C_R$ (right column) for the three parametrizations. As a reference, we have marked the Casimir scaling expectation, given by $C_R = N_C / C_F = 2.25$, in the latter. The prior distributions are also presented (open histograms in Fig.~\ref{fig:mel}), in order to show the unbiased constraint that results from the Bayesian inference.

A clear separation is seen between the mean energy lost by a quark-initiated jets and gluon-initiated jets, for the normal and gamma distributions, with the gluon-initiated jets loosing more energy on average as expected, see Sec. \ref{sec:theory-poisson}. Surprisingly, no apparent difference is seen between the energy loss for the different parton-initiated jets when the log-normal parametrization is used.  These differences are even more pronounced when evaluating the color ratio factor. The three parametrizations give the central values $C_R \approx 3.5$ (normal), $C_R \approx 1$ (log-normal) and $C_R \approx 2.25$ (gamma).

This result is concerning because, while all parametrizations have been carefully adjusted to provide a satisfactory description of the experimental data, the underlying physical properties they represent\,\textemdash\, if we are justified in calling them that\,\textemdash\, differ significantly. This raises questions about whether the parametrizations truly capture the essential physics or if they are merely effective descriptions that fit the data without reflecting the actual mechanisms at play.

\begin{figure*}
    \centering
    \includegraphics[width=0.9\textwidth]{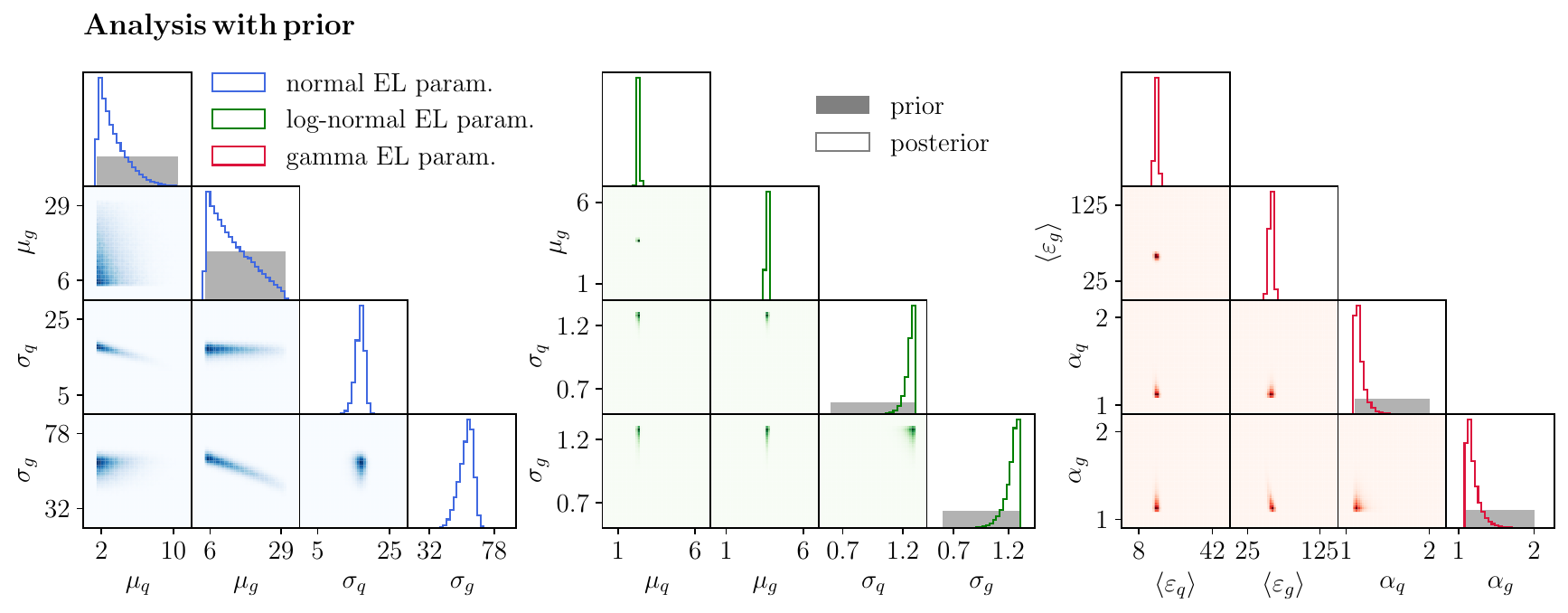}
    \caption{Same as above, when prior constraining is used.}
    \label{fig:corner_wprior}
\end{figure*}

\subsection{Restricting the prior range}
\label{sec:prior-constraint}

We attribute this inconsistency of extracting $C_R$ for the different parametrizations, as seen in Fig.~\ref{fig:mel}, to the shape of the energy-loss parametrization distributions. The steeply falling jet spectrum introduces a bias that makes it difficult to constrain the tails of the distributions. This bias affects more severely the constraining of the mean for distributions with a heavy tail, which is the case of the log-normal distribution, than distributions with a more compact support, such as the normal or gamma distributions. In Sec.~\ref{sec:parametrizations}, we have discussed the main features of the three parametrizations we are using in this work. The quantitative dependence of the mean on the shape of the distribution can be assessed by the mean-mode ratio $\langle\varepsilon\rangle/\varepsilon_{\rm max}$ of the mean energy-loss. For the log-normal distribution this ratio, see Eq.~\eqref{eq:mean_max_lognormal}, grows exponentially with the standard deviation $\sigma$. For the normal distribution the mean is always close to the peak of the distribution by construction. For the gamma distribution, as long as $\alpha \neq 1$, the relation is also quite tight.

To mitigate this bias, we can manually impose a limit on the relative distance between the mean and the mode of each energy-loss parametrization distribution. We get this limit by imposing a physical constraint on the allowed values of $\alpha_s$, which relates to the mean-mode ratio of the energy loss via Eq. \eqref{eq:mean_mode_theory}. For weakly-coupled scenarios we consider $0.1 < \alpha_s< 0.5$. This limit in $\alpha_S$ translates into a limit on the range of the prior distribution used for Bayesian inference, summarized in Tab.~\ref{tab:restriction}. We impose these limits into our analysis by introducing them as priors over the model parameters.

\begin{table}[]
    \renewcommand{\arraystretch}{1.2}
    \centering
    \begin{tabular}{c|c|c}
         parametrization & restriction on & range \\
         \hline
         normal & $\xi = \mu/\sigma$ & [0.08, 0.51] \\
         log-normal & $\sigma$ & [0.68, 1.24] \\
         gamma & $\alpha$ & [1.11, 2]
    \end{tabular}
    \caption{Range for parameter sampling after constraint is imposed.}
    \label{tab:restriction}
\end{table}

Figure~\ref{fig:corner_wprior} shows the resulting posterior distributions when the restricted prior described above is used. The posterior distributions are far more constrained compared with the posterior distribution obtained when no physically informed prior is used, see Fig.~\ref{fig:corner_global}. Apart for the normal distribution, which we will come back to below, there is very little correlation between the observables. These results demonstrate that our streamlined parametrization of jet energy loss, which, for example, neglects significant $\pT$ variation over the measured ranges, successfully captures the key features of the data.

\begin{figure*}
    \centering
    \includegraphics[width=\textwidth]{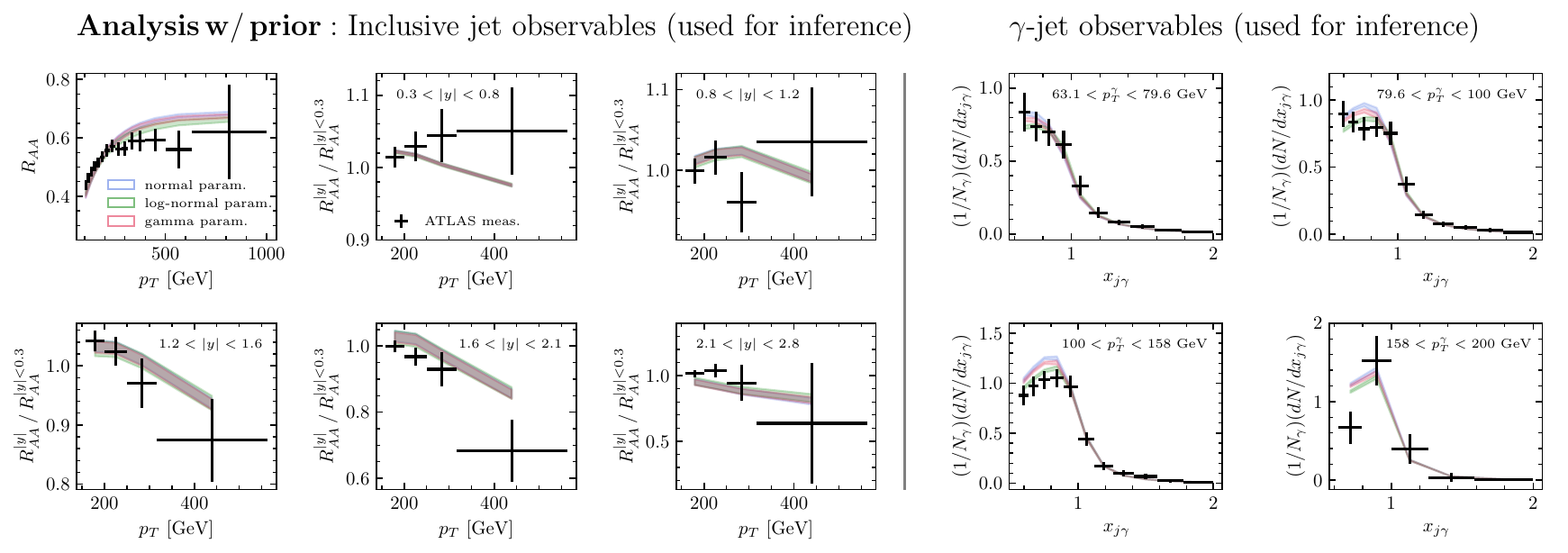}
    \caption{Same as above, when prior constraining is used.}
    \label{fig:fit_pred_c_wprior}
\end{figure*}
At the same time the posterior predictive for the observables in study, shown in Fig.~\ref{fig:fit_pred_c_wprior}, are more consistent between the three energy-loss parametrizations. The solutions of the inference have also decisively converged, resulting in a very narrow band. Translating the information in the posterior distributions in Fig.~\ref{fig:corner_wprior} into $\alpha_s$, we conclude that the data tends to prefer very small values of $\alpha_s$. For example, for the gamma distribution $\alpha_s \sim \alpha-1 \ll 1$ which is observed both for quark- and gluon-initiated jets in Fig.~\ref{fig:corner_wprior} (left panel).

\begin{figure}
    \centering
    \includegraphics[width=0.9\columnwidth]{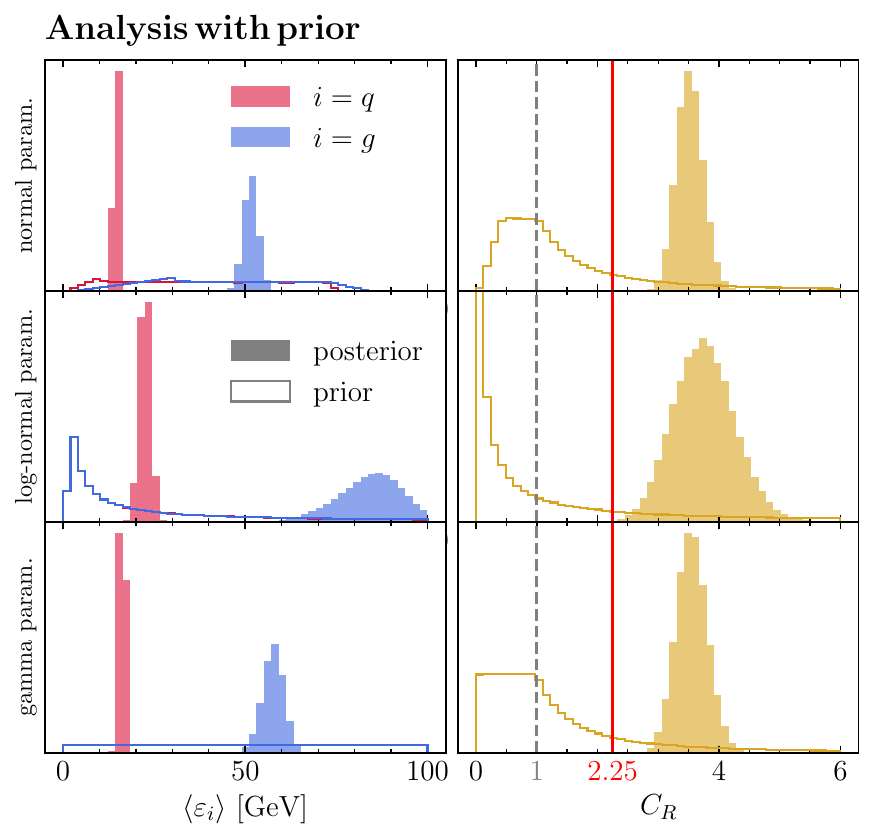}
    \caption{Same as above, when prior constraining is used.}
    \label{fig:mel_wprior}
\end{figure}
The introduced prior also highly constrains the mean-energy loss and the respective color ratio, shown in Fig.~\ref{fig:mel_wprior}. A first observation is that the sampled mean energy losses are quite narrowly distributed, especially for the energy loss off quark-initiated jets. Secondly, we notice that now a consistent color ratio, with a relatively small variance, is achieved for the three energy-loss parametrizations. The central value of the extracted color ration of energy loss is $C_R \approx 3.5$. This can be interpreted as an anomalous behavior of energy loss on the Casimir ratio, i.e.
\begin{equation}
    \label{eq:super-casimir}
    C_R = \left(\frac{N_c}{C_F} \right)^{1+\gamma} \,,
\end{equation}
where the extracted value is $\gamma \approx 0.5$. We refer to this feature as \textit{super-}Casimir scaling.

\begin{figure}
    \centering
    \includegraphics[width=0.9\columnwidth]{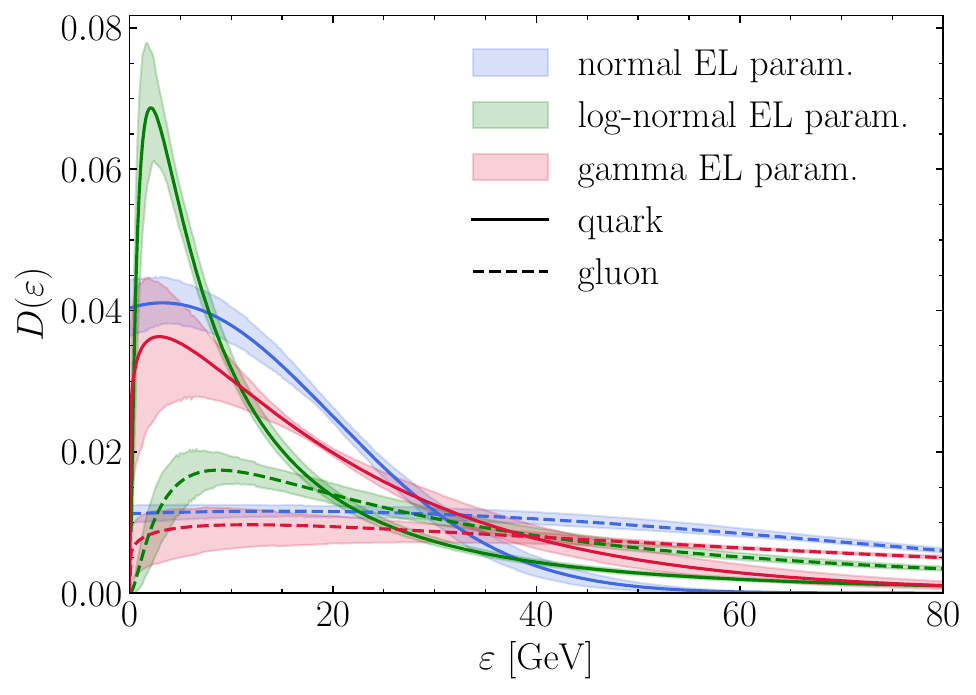}
    \caption{Posterior mean and 90\% HDI of the energy loss distributions from global inference with restricted prior, for the three parametrizations in study.}
    \label{fig:el_dist}
\end{figure}
Finally, we plot the posterior mean and 90\% HDI of the energy loss distributions $D_i(\varepsilon)$ obtained for the three energy loss parametrizations in Fig.~\ref{fig:el_dist}. The spread between the actual parametrizations is considerable considering the tight constraints imposed by the experimental data, see Fig.~\ref{fig:fit_pred_c_wprior}. This can be attributed to jet suppression bias and demonstrates how this effect imposes a strong limitation on extracting physical information about jet energy loss from existing experimental data \,\textemdash\, at least the class of jet observables considered in this work.

The second feature of the extracted energy loss distributions is the relatively fat tails and the almost complete absence of a peak of the distributions, see especially the shape of $D_g(\varepsilon)$ for the normal and gamma parametrizations in Fig.~\ref{fig:el_dist}. Note that the shown energy loss distributions do not necessarily resemble the original form of the underlying parametrization, since the plotted curves correspond to the mean of the posteriors of $D_i(\varepsilon)$ at each value of $\varepsilon$. The absence of a peak could potentially be related to the lack of a characteristic energy scale, which may stem from the medium's energy density and geometry, due to the combination of confounding effects (such as energy loss versus medium response) and severe fluctuations (such as path-length fluctuations or intrinsic fluctuations of energy loss processes).

The fat tails are qualitatively similar to a Pareto-type distribution, which can also indicate that rare processes, involving large energy loss, can be responsible for the observed effects\,\textemdash\, quite contrary to the theoretical expectations expounded in Sec.~\ref{sec:theory}. This serves to remind us that we should take the extracted distributions in Fig.~\ref{fig:el_dist} with a grain of salt as they are simplified representations of complex jet-by-jet dynamics and, at the same time, reconstructed through the lens of biased observables. 

\subsection{Comment on the use of other data}
\label{sec:other-data}

The global analysis above has been performed using a specific set of available experimental data. There is however also available data at the LHC on inclusive jet suppression from CMS \cite{CMS:2011iwn,CMS:2016uxf} and for photon-tagged jets in a wide range of $\pT$, with $\pTg > 40$ GeV and $\pTj > 30$ GeV \cite{CMS:2017ehl}. The ALICE collaboration has also measured charged jets \cite{ALICE:2015mjv,ALICE:2015mdb} at slightly lower $\pT < 200$ GeV than CMS and ATLAS. 

There is a slight tension on the level of overall suppression between CMS \cite{CMS:2016uxf} and ATLAS \cite{ATLAS:2018gwx} in the moderate-to-high $\pT$ range, $100 \text{ GeV}< \pT < 300 \text{ GeV}$, at mid-rapidity where the most precise data currently exists. In order to take advantage of the tight constraints from either of the data-sets we finally opted for sticking with the ATLAS $R_{AA}$ data for our final analysis. It would be interesting to redo the present analysis by including, in a proper fashion, both the CMS $R_{AA}$ and the ALICE data at lower $\pT$ to fully explore the $\pT$ dependence of energy loss. For further applications, the $R$-dependent jet suppression could also be considered \cite{CMS:2021vui,ATLAS:2023hso}.

The photon-tagged $R_{AA}$ \cite{CMS:2017ehl}, on the other hand, poses an interesting challenge to test our predictions due to the interesting behavior of the relatively stable and large contribution of quark-initiated jets, compared to a conventional inclusive jet sample, see Fig.~\ref{fig:qrk_frac}. It therefore deserves a special discussion.

\begin{figure}
    \centering
    \includegraphics[width=0.9\columnwidth]{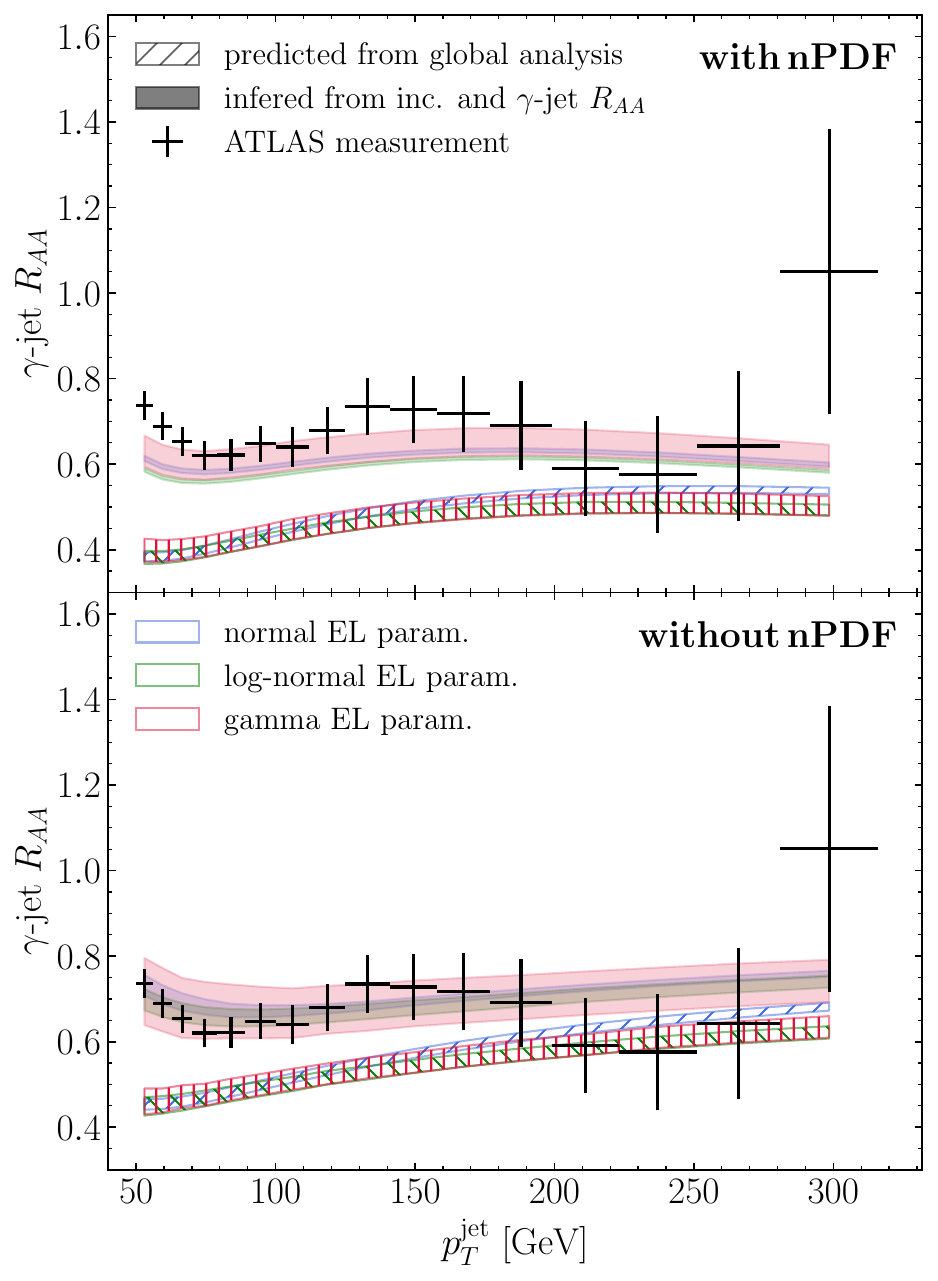}
    \caption{90\% HDI of the posterior predictive distribution for the photon-tagged jet $R_{AA}$, for the three energy-loss parametrizations. In solid color is the result when only the inclusive $R_{AA}$ and the photon-tagged $R_{AA}$ are used for inference while in open color is the prediction from an inference that used all the observables but the photon-tagged $R_{AA}$. Upper panel: with nPDF. Lower panel: nPDF is not used here.}
    \label{fig:gamma_raa}
\end{figure}
We have addressed the photon-tagged jet data in our analysis in two ways:
\begin{itemize}
    \item by predicting the observable based on the inference from the global analysis of remaining experimental data, and
    \item by including it into a new analysis.
\end{itemize}
In this context we have also explored the impact of nuclear PDFs, especially for quark-initiated jets at high-$\pT$ \footnote{These effects are remnants of the EMC effect and is parameterized from nuclear DIS data at low-$Q$. Since the quark non-singlet decouples from the gluons, this feature is protected by DGLAP evolution.}. For more details, see App.~\ref{sec:photon-tagged}.

We show the results of our analysis using EPS09 nPDFs (upper panel) and conventional proton PDFs (lower panel) in Fig.~\ref{fig:gamma_raa}. All three parametrizations have been employed showing complete consistency \footnote{We have also not observed any particular effect of removing the nPDFs from our ``standard'' global analysis.}. If we  predict the photon-tagged suppression based on the previous global analysis with nPDFs, we over-estimate the suppression by a factor of 50\% at $\pT \approx 100$ GeV (textured bands, upper panel). This is reduced to only a 20\% over-estimation when we turn off nPDFs and a much better description at high-$\pT$ (textured bands, lower panel).

Interestingly, a good agreement with the data is achieved when inference is performed using only the inclusive and photon-tagged $R_{AA}$ (colored bands in upper and lower panels of Fig.~\ref{fig:gamma_raa}). Leaving out the rapidity dependence of the inclusive $R_{AA}$ and the per-photon $\gamma$-jet yields of our analysis, as in, e.g., \cite{Zhang:2023oid}, significantly improves the agreement with the data, as seen in Fig.~\ref{fig:gamma_raa} (colored bands in upper panel). The absence of nuclear effects on PDFs, also not included in \cite{Zhang:2023oid}, further improves the agreement with the data. However, we notice a significant worsening of the description of inclusive jet $R_{AA}$ at forward rapidities when such an analysis is performed, which was much less severe in the absence of nuclear effects on PDFs. For more details see App.~\ref{sec:photon-tagged}.

This indicates that there is a certain tension between the inclusive jet and photon-tagged jet data. It would be interesting to see whether the photon-tagged jet data has a particular sensitivity to nuclear effects on the PDFs. We believe that a recalibration of the overall inclusive jet suppression at mid-rapidity, by including more data \cite{CMS:2016uxf}, could also help in achieving consistency. We leave these more detailed analyses for future work.

\section{Conclusions}
\label{sec:conclusions}

Jet modifications in the dense nuclear environment created during heavy-ion collisions at the LHC serve as valuable tools for probing the quark-gluon plasma. Beyond the ambitious goal of extracting transport properties, one can pose more fundamental questions. For example, what are the generic features of jet suppression when compared to well-established baselines from $pp$ jet physics? Additionally, is the medium sensitive to the partonic composition of the developing QCD jet? We have attempted to answer these questions using a fully data-driven extraction of the so-called energy loss distribution responsible for jet energy loss with minimal assumptions from theory. This has been achieved using Bayesian inference. 

In contrast to other Bayesian analyses of soft and hard observables in heavy-ion collisions, we have deliberately avoided constraining parameters within a specific model that aims to describe relevant aspects of the observables. Instead, we employ a range of flexible distributions, that satisfy appropriate criteria, to gauge the constraining power of existing experimental data. In order to separate and constrain the color dependence of energy loss, we have analyzed experimental data with different fractions of quark- and gluon-initiated jets in the baseline prediction. An extensive statistical analysis have been performed in order to scrutinize and interpret our numerical results.

Our main conclusions can be summarized in the following points.
\begin{itemize}
    \item Bias due to the steeply falling nature of the underlying jet spectrum is strongly affecting our ability to constrain the energy loss distribution at all relevant scales. This holds both for observables based on the inclusive jet spectrum as well as for the coincidence photon-jet measurements, although the problem is less severe in the latter case. This has been established through closure tests and the global analysis of experimental data.
    \item We have established evidence of \textit{universality}, i.e. that the extracted energy loss distribution is applicable to different baseline processes under the factorization in Eq.~\eqref{eq:fact}.
    \item We have established evidence of color charge dependence of energy loss, see Fig.~\ref{fig:mel_wprior}. This was summarized in Eq.~\eqref{eq:super-casimir}.
    \item We have not been able to extract a characteristic energy scale associated with medium interactions in a finite-size medium.
    \item We have established consistency between the analyzed data-sets through a leave-one-out analysis in Sec.~\ref{sec:leave-one-out}. We have also extensively analyzed photon-tagged jet data in Sec.~\ref{sec:other-data} and shed light on its sensitivity to nPDF effects.
\end{itemize}

It seems pertinent to continue to look for strategies to mitigate the inherent bias of jet observables in order to allow experimental data to put more stringent constraints on theoretical descriptions and modeling of jet quenching phenomena in heavy-ion collisions.

\begin{acknowledgments}
    The authors would like to thank Adam Takacs, Raymond Ehlers for constructive discussions, and Yi Chen for help with setting up the inference framework.
\end{acknowledgments}

\appendix

\section{Event generation baseline and fits}
\label{sec:spectra_fits}
\begin{figure*}
    \centering
    \includegraphics[width=.85\textwidth]{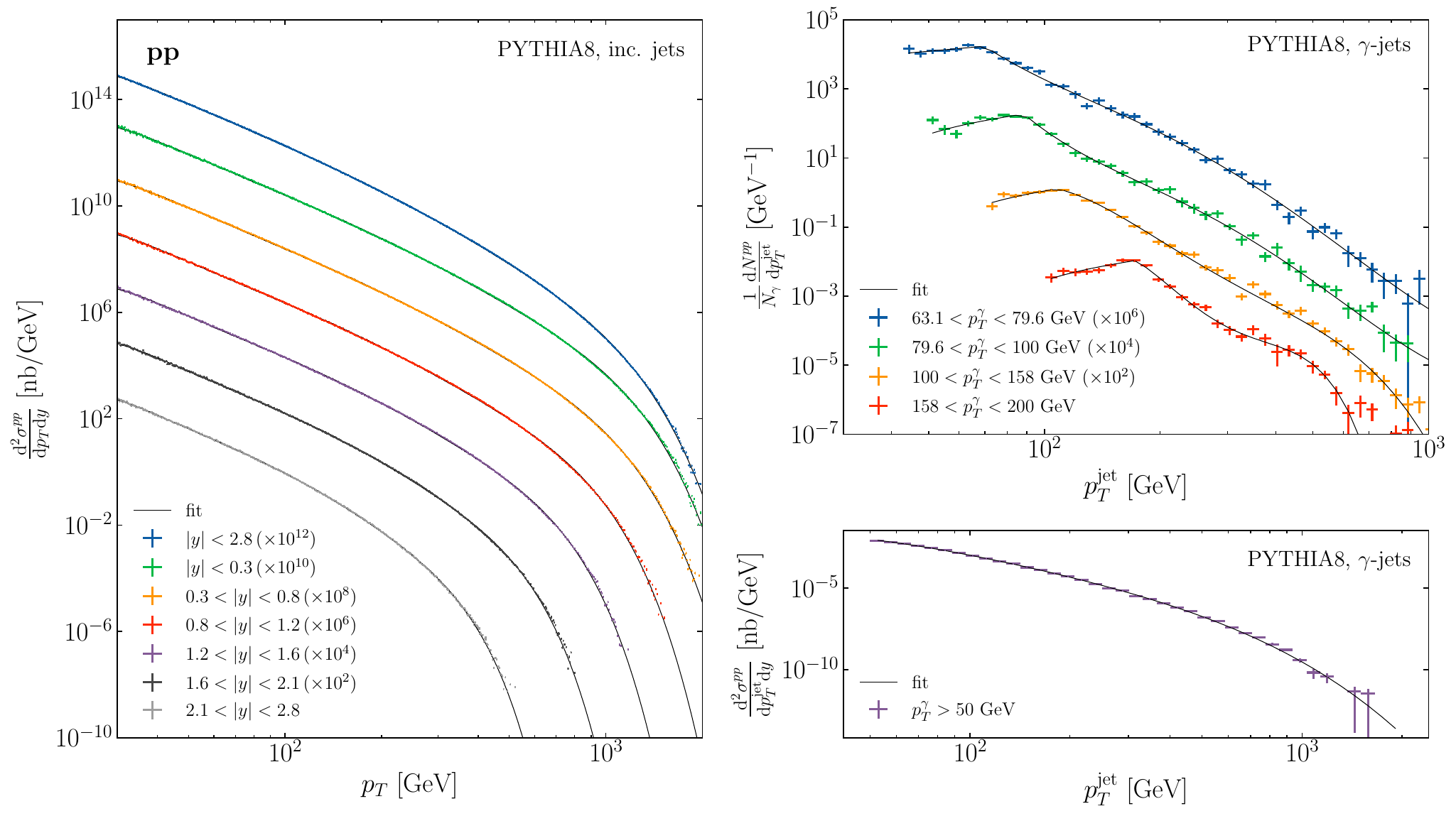}
    \caption{Fits to the PYTHIA8 $pp$ collisions (with standard PDFs) sampled inclusive jet cross-section (left panel), and on the photon-tagged jet yield (upper left panel) and cross-section (lower left panel), as functions of $p_T$.}
    \label{fig:pp_fit}
\end{figure*}

\begin{figure*}
    \centering
    \includegraphics[width=.85\textwidth]{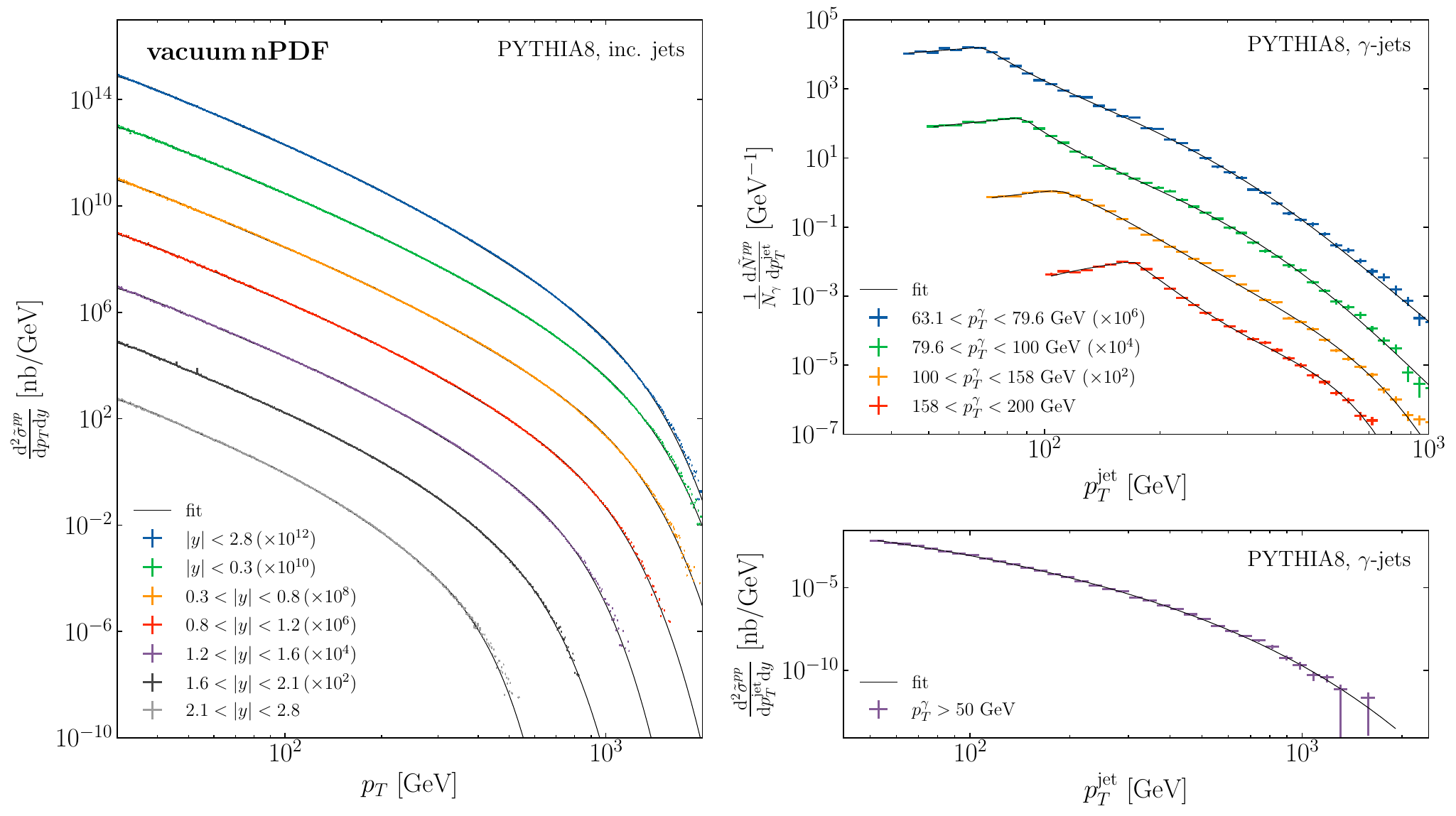}
    \caption{Fits to the PYTHIA8 nucleus-nucleus collisions sampled inclusive jet cross-section (left panel), and on the photon-tagged jet yield (upper left panel) and cross-section (lower left panel), as functions of $p_T$.}
    \label{fig:AA_fit}
\end{figure*}

In this appendix, we provide detailed information about the fits of the baselines used in the paper. These results are summarized in Figs.~\ref{fig:pp_fit} and \ref{fig:AA_fit}. We plot three types of observables:
\begin{itemize}
    \item inclusive jet spectrum for different ranges in jet rapidity $y$,
    \item per-photon jet yield for different photon $\pTg$ bins, and
    \item photon-tagged jet spectrum with $\pTg > 50$ GeV. 
\end{itemize}
The generated baseline data for $pp$ collisions, i.e. using conventional PDFs, in Fig.~\ref{fig:pp_fit} and nucleus-nucleus collisions, i.e. with isospin effects and nuclear PDFs encoded in the EPS09 PDF set, in Fig.~\ref{fig:AA_fit}, are plotted as points (with error bars), while the fits are plotted with continuous lines. The fits for inclusive and photon-tagged jet spectra are based on the form in Eq.~\eqref{eq:fall_spec}, while the fits for the per-photon yield is based on Eq.~\eqref{eq:balance_spec}. Overall we achieve an excellent description of the baseline data with the proposed fitting procedure.

\section{Study on $\gamma$-jet $R_{AA}$ and nPDF effects}
\label{sec:photon-tagged}
\begin{figure*}
    \centering
    \includegraphics[width=\textwidth]{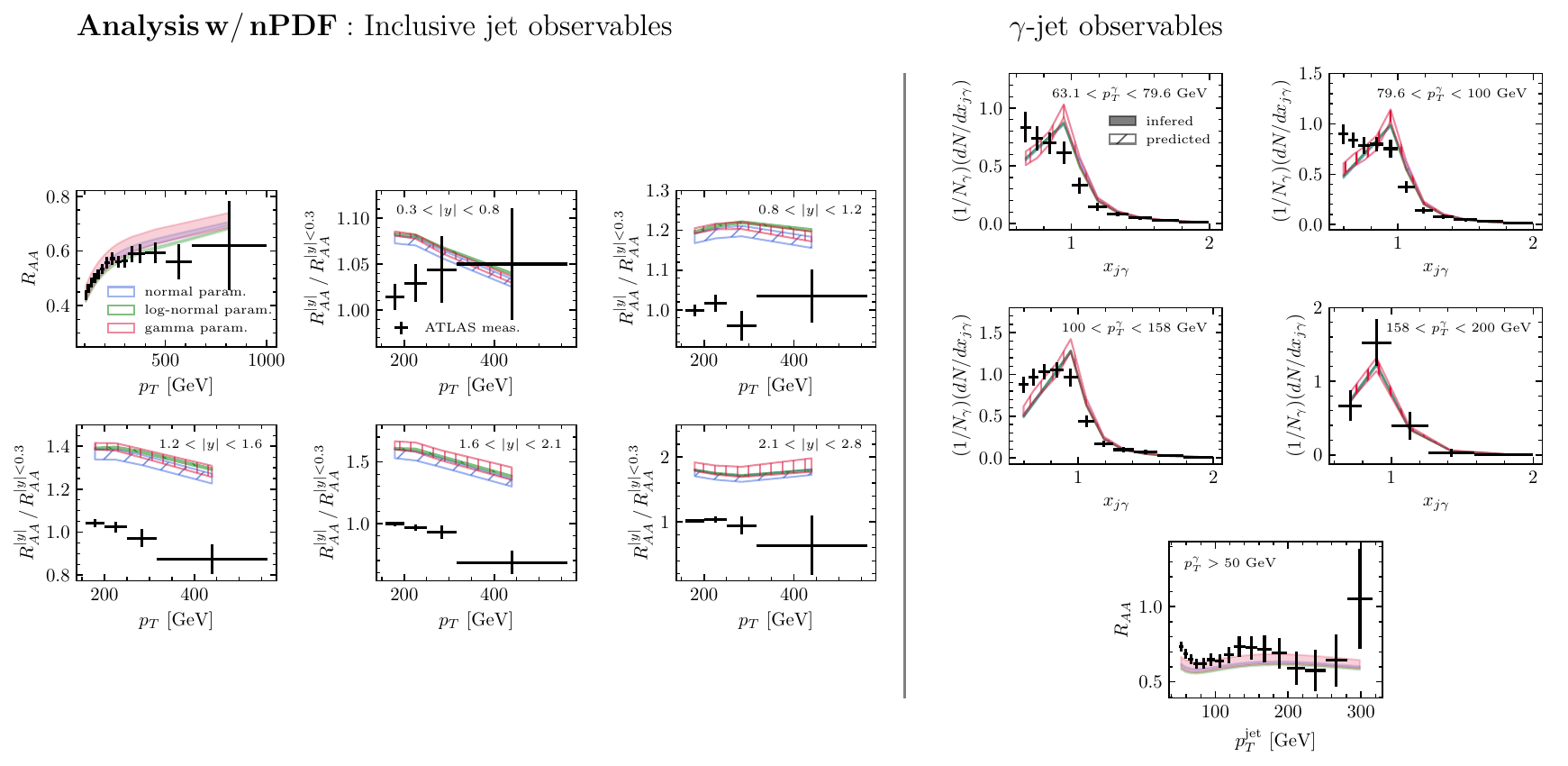}\\
    \includegraphics[width=\textwidth]{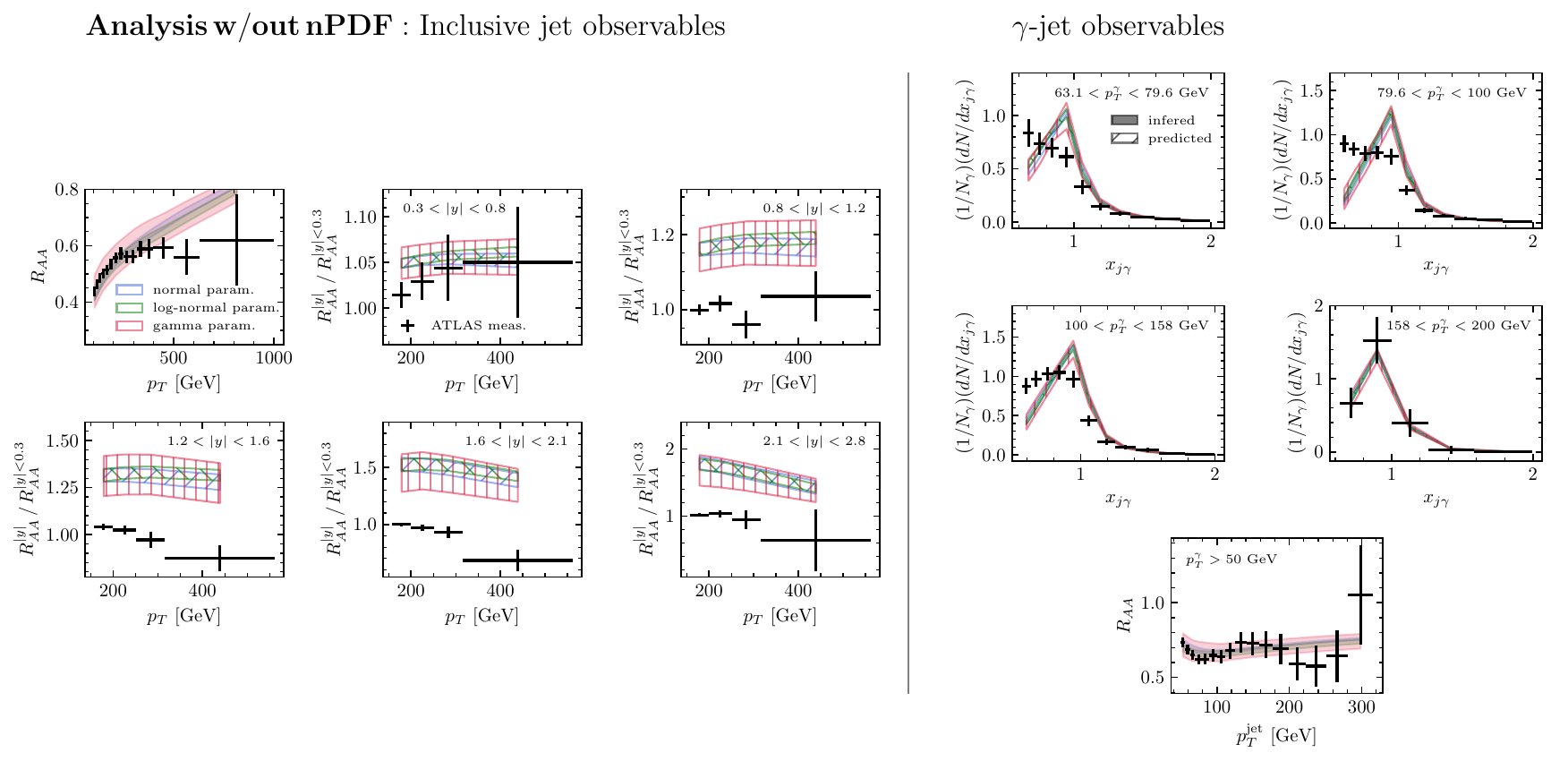}
    \caption{90\% HDI of the posterior predicative distributions obtained for all the observables in the study, when Bayesian inference is done using only the inclusive and $\gamma$-jet $R_{AA}$. A solid color is used in the observables used in for inference (inclusive and photon-tagged jets), while textured ones are predicted (forward-rapidity inclusive jets and per-photon jet yields). The analysis in the upper panels include nPDF effects, while the lower panels neglect nPDF effects.}
    \label{fig:fit_pred_h}
\end{figure*}

In this appendix, we provide further details on the results of the Bayesian analyses that include the photon-tagged jet $R_{AA}$ data from \cite{ATLAS:2023iad}. Bayesian inference was performed using the inclusive and photon-tagged jet $R_{AA}$. The obtained posterior distributions were then used to predict the remaining data, i.e. rapidity dependent relative $R_{AA}$ and per-photon $\gamma$-jet yield for different $\pTg$ bins. Figure~\ref{fig:fit_pred_h} shows the 90\% HDI of the obtained posterior predictive distributions, both for the data points used and not used for inference, together with the respective measured data. The upper panel of Fig.~\ref{fig:fit_pred_h} shows the results when the analysis was done using nPDF effects, and the lower panel when the analysis was done when neglecting nPDF effects, i.e. using ordinary proton PDFs for the heavy-ion baseline.

We see that by only using the inclusive and $\gamma$-jet $R_{AA}$ for inference, the mid-rapidity observables are successfully described by the computed posterior predictive distributions. However, this comes with a significant worsening of the description of the inclusive $R_{AA}$ at forward rapidity. This incorrect description is more pronounced in the presence of nPDF effects, see the upper panel in Fig.~\ref{fig:fit_pred_h}, because the 90\% HDI bands are significantly narrower. Regarding the per-photon jet yield at $\xjg \approx 1$ (right panels of Fig.~\ref{fig:fit_pred_h}), we also see a decrease in the agreement between the posterior predictive and the measured data, for both analyses with and without nPDF effects. More worrying, the absence of nPDFs significantly deteriorate the description of high-$\pT$ behavior for inclusive jet $R_{AA}$ at mid-rapidity.

In Fig.~\ref{fig:nPDF_effect}, we presented the nPDF effect on the inclusive jet cross-section, and on the photon-tagged jet yield, as a function of the jet $\pT$. The seen nPDF effect is responsible for a suppression enhancement of high-$\pT$ inclusive jets, across rapidity. This suppression enhancement can be clearly seen in Fig.~\ref{fig:fit_pred_h}, by comparing the hight-$\pT$ region of the inclusive $R_{AA}$ with (upper panel), and without (lower panel) the use of nPDFs. The nPDF effect has therefore an important contribution to the correct description of the high-$\pT$ region of the inclusive $R_{AA}$. Furthermore, the nPDFs are also important for the $\gamma$-jet $R_{AA}$ since the nPDF effect accounts already for about 20\% of the overall jet suppression, as can be seen in Fig.~\ref{fig:nPDF_effect}. Therefore, not using of nPDFs would have to be compensated by enforcing stronger final-state jet quenching \,\textemdash\, mainly of quark-initiated jets.

The presence of the nPDF ultimately affect the extracted mean energy loss of quark- and gluon-initiated jets and it would be desirable to constrain it independently using other means, e.g. in proton-nucleus collisions.

\bibliography{references}

\end{document}